\documentclass[journal=jpcbfk,manuscript=article]{achemso}
\usepackage{longtable}
\usepackage{multirow}
\usepackage{amsmath}
\usepackage[usenames,dvipsnames]{color}
\usepackage{soul}
\usepackage{threeparttable}
\usepackage{xr}
\usepackage{graphicx}
\usepackage{amsmath,amssymb}
\usepackage{caption}
\usepackage{color}
\usepackage{xcolor}
\usepackage{dcolumn}
\usepackage{bm}
\usepackage{float}
\usepackage{subcaption}
\usepackage{textcomp}
\usepackage{url}
\usepackage[normalem]{ulem}

\usepackage{microtype}
\usepackage{etoolbox}

\usepackage[unicode=true,
            bookmarks=true,
            bookmarksnumbered=true,
            bookmarksopen=true,
            bookmarksopenlevel=2,
            breaklinks=false,
            pdfborder={0 0 1},
            backref=false,
            colorlinks=true,
            hidelinks
]{hyperref}

\author{Meenu Upadhyay} \affiliation[University of Basel]{Department
  of Chemistry, University of Basel, Klingelbergstrasse 80, CH-4056
  Basel, Switzerland.}

\author{Silvan Käser} \affiliation[University of Basel]{Department of
  Chemistry, University of Basel, Klingelbergstrasse 80, CH-4056
  Basel, Switzerland.}\altaffiliation{Present Address: Roche Pharma
  Research and Early Development, Pharmaceutical Sciences, Roche
  Innovation Center Basel, F. Hoffmann-La Roche Ltd, Basel,
  Switzerland}

\author{Jayakrushna Sahoo} \affiliation[Université de
  Montpellier]{Laboratoire Univers et Particules de Montpellier,
  Université de Montpellier, UMR-CNRS 5299, 34095 Montpellier Cedex,
  France.}

\author{Yohann Scribano} \affiliation[Université de
  Montpellier]{Laboratoire Univers et Particules de Montpellier,
  Université de Montpellier, UMR-CNRS 5299, 34095 Montpellier Cedex,
  France.}

\author{Markus Meuwly} \affiliation[University of Basel]{Department of
  Chemistry, University of Basel, Klingelbergstrasse 80, CH-4056
  Basel, Switzerland.}  \email{m.meuwly@unibas.ch}

\title{Reaction Dynamics of the H + HeH$^+$ $\rightarrow$ He + H$_2^+$
  System}
\begin{document}

\date{\today}

\begin{abstract}
The reaction dynamics for the H + HeH$^+$ $\rightarrow$ He + H$_2^+$
reaction in its electronic ground state is investigated using two
different representations of the potential energy surface (PES). The
first uses a combined kernel and neural network representation of
UCCSD(T) reference data whereas the second is a corrected PES (cR-PES)
that eliminates an artificial barrier in the entrance channel
appearing in its initial expansion based on full configuration
interaction reference data. Despite the differences between the two
PESs, both yield $k_{v=0,j=0} \approx 2 \times 10^{-9}$
cm$^3$/molecule/s at $T = 10$ K which is consistent with a
$T-$independent Langevin rate $k_{\rm L} = 2.1 \times 10^{-9}$
cm$^3$/molecule/s but considerably larger than the only experimentally
reported value $k_{\rm ICR} = (9.1 \pm 2.5) \times 10^{-10}$
cm$^3$/molecule/s from ion cyclotron resonance experiments. Similarly,
branching ratios for the reaction outcomes are comparable for the two
PESs. However, when analysing less averaged properties such as initial
state-selected $T-$dependent rate coefficients and final vibrational
states of the H$_2^+$ product for low temperatures, the differences in
the two PESs manifest themselves in the observables. Thus, depending
on the property analyzed, accurate and globally valid representations
of the PES are required, whereas more approximate and empirical
construction schemes can be followed for state-averaged observables.
\end{abstract}

\section{Introduction}
The HeH$^+$ ion plays a central role in the chemical evolution of the
universe. This ion was probably the first molecule to be formed and it
plays an important role in the formation of H$_2$ in the early
universe.\cite{geppert:2013,courtney:2021} Direct detection of HeH$^+$
in the interstellar space was possible in the planetary nebulae NGC
7027\cite{stutzki:2019,neufeld:2020} and resolved a long-standing
dilemma for astronomy. In the laboratory, HeH$^+$ was generated as
early as 1925 by electron impact of a mixture containing H$_2$ and
He.\cite{hogness:1925}\\

\noindent
One of the destruction processes for HeH$^+$ is its interaction with
atomic hydrogen which generates He + H$_2^+$.\cite{geppert:2013} To
investigate and understand the spectroscopy and reaction dynamics of
small, reactive species in the interstellar
environment\cite{tielens:2013} quantum nuclear dynamics and/or
quasi-classical trajectory (QCT) simulations can be used. Both require
sufficiently precise descriptions and representations of the inter- and intramolecular
interactions by way of a global potential energy surface (PES) because
the on-the-fly evaluation of energies and forces at the required
levels of theory, such as coupled cluster with perturbative triples
(CCSD(T)), multi reference and full configuration interaction (MRCI
and FCI) calculations is not possible. Hence, for reliable and
meaningful results to compare with and eventually predict experimental
observables, high-level reference data for the interaction potential,
accurate methods for representing it, and reliable methods for
treating the nuclear dynamics using such a PES are required.\\

\noindent
Despite the accuracy of the underlying electronic structure
calculations, the represented PESs must still be scrutinized in view
of their overall accuracy and shape. One example for this is a ``reef"
structure that was found for the O$(^3{\rm P})$ +
O$_2(^3\Sigma_g^{-})$ recombination reaction to form
O$_3$.\cite{fleurat:2003,babikov:2003,ayouz:2013,Dawes:2013,Li:2014,tyuterev:2014}
Although high-level electronic structure methods at different levels
of theory, including MRCI calculations, find the
``reef"\cite{pack:2002,schinke:2004,holka:2010}, removing this feature
yielded improved agreement between computed and experimental
observations for thermal rate coefficients.\cite{fleurat:2003} Also,
the absence of a ``reef" was directly linked to a negative temperature
dependence of the thermal rate $k(T)$ from wavepacket calculations
which is consistent with observations.\cite{Li:2014} Similarly, for
the N($^4$S) + NO(X$^2\Pi$) $\rightarrow$ O($^3$P) + N$_2$(X$^1
\Sigma^+_g$) reaction which is relevant in hypersonics and atmospheric
chemistry, a spurious barrier at long range led to ``turnover" of the
thermal rate $k(T)$ for $T \lesssim 50$ K, inconsistent with
experiment.\cite{MM.n2o:2023} The PES, originally developed for
high-energy reaction dynamics, revealed a spurious barrier (38 K) in
the entrance channel when reactions at low temperatures were
investigated.\cite{MM.n2o:2020} The origin of this spurious feature
was that the underlying grid on which the reference data were
calculated covered a range that was too narrow.\\

\noindent
Another example for a PES exhibiting undesired features in the
entrance channel are the polynomial representations of MRCI- and
FCI-based reference calculations\cite{ramachandran:2009} to
investigate the H + HeH$^+$ $\rightarrow$ He + H$_2^+$ reaction. This
process is expected to be barrierless but computed integral cross
sections abruptly decrease for low collision energies ($< 10$ meV) due
to small but unphysical barriers between 0.66 and 4.8 meV depending on
the level of theory.\cite{sahoo.refined.2024} By combining 2-body
interactions from a previous full configuration interaction
study\cite{MM.heh2:2019} including the physical long-range behaviour
with 3-body interactions from the original
PESs\cite{ramachandran:2009} the spurious barrier was removed and the
integral cross sections remain finite down to the lowest collision
energies.\cite{sahoo.refined.2024} The validity of the FCI level of
theory for the He--H$_2^+$ channel\cite{MM.heh2:2019} was
independently and successfully established from quantum wavepacket
calculations to interpret Feshbach resonance measurements and final
adjustment to experimental data through
PES-morphing.\cite{MM.heh2:2023,MM.morphing:2024,MM.morphing:1999}\\

\noindent
The present work uses and evaluates an accurate mixed kernel/neural
network (NN) representation of a fully reactive PES for the [HHHe]$^+$
system\cite{mm.kernn:2024} The initial state-selected $T-$dependent
rate coefficients $k_{v=0,j \in [0,5]}(T)$ up to 1000 K are
characterized from classical and quantum nuclear dynamics simulations
using this KerNN PES and a recently corrected PES
(cR-PES)\cite{sahoo.refined.2024} of the RFCI8
PES.\cite{ramachandran:2009} First, the potential energy surfaces are
presented and validated, followed by the results on thermal rates and
final state distributions. Next, the results are discussed in a
broader context and conclusions are drawn.\\

\section{Results}
First, the properties of the new KerNN representation of the reactive
PES are described and compared with the cR-PES for the H + HeH$^+$
$\rightarrow$ He + H$_2^+$ reaction. Next, $T$-dependent rate
coefficients determined from QCT and TIQM simulations are discussed,
followed by an analysis of product state branching ratios and final
state distributions. Finally, the findings are considered in a broader
context with particular emphasis on the accuracy of the results and
their implications.

\subsection{Characterization of the PESs}
For the KerNN PES trained on UCCSD(T)/aug-cc-pV5Z reference data the
reaction considered is exothermic with the reactant H$_{\rm A}$ +
HeH$_{\rm B}^+$ at --2.040747~eV, the (stabilized) collinear
intermediate [HeH$_{\rm A}$H$_{\rm B}$]$^+$ at --3.131270~eV, and the
dissociated product He + H$_2^+$ at --2.792732~eV relative to the
separated atoms He + H + H$^+$. The reaction is expected to feature no
barrier in the entrance channel with He + H$_2^+$ as the main
product. However, other processes are possible, including atom
exchange H$_{\rm A}$ + HeH$_{\rm B}^+$ $\longrightarrow$ H$_{\rm B}$ +
HeH$_{\rm A}^+$, and elastic and inelastic collisions H$_{\rm A}$ +
HeH$_{\rm B}^+ (v, j)$ $\longrightarrow$ H$_{\rm A}$ + HeH$_{\rm B}^+
(v/v', j/j')$.\\

\begin{figure}[H]
\centering
\includegraphics[width=0.7\textwidth]{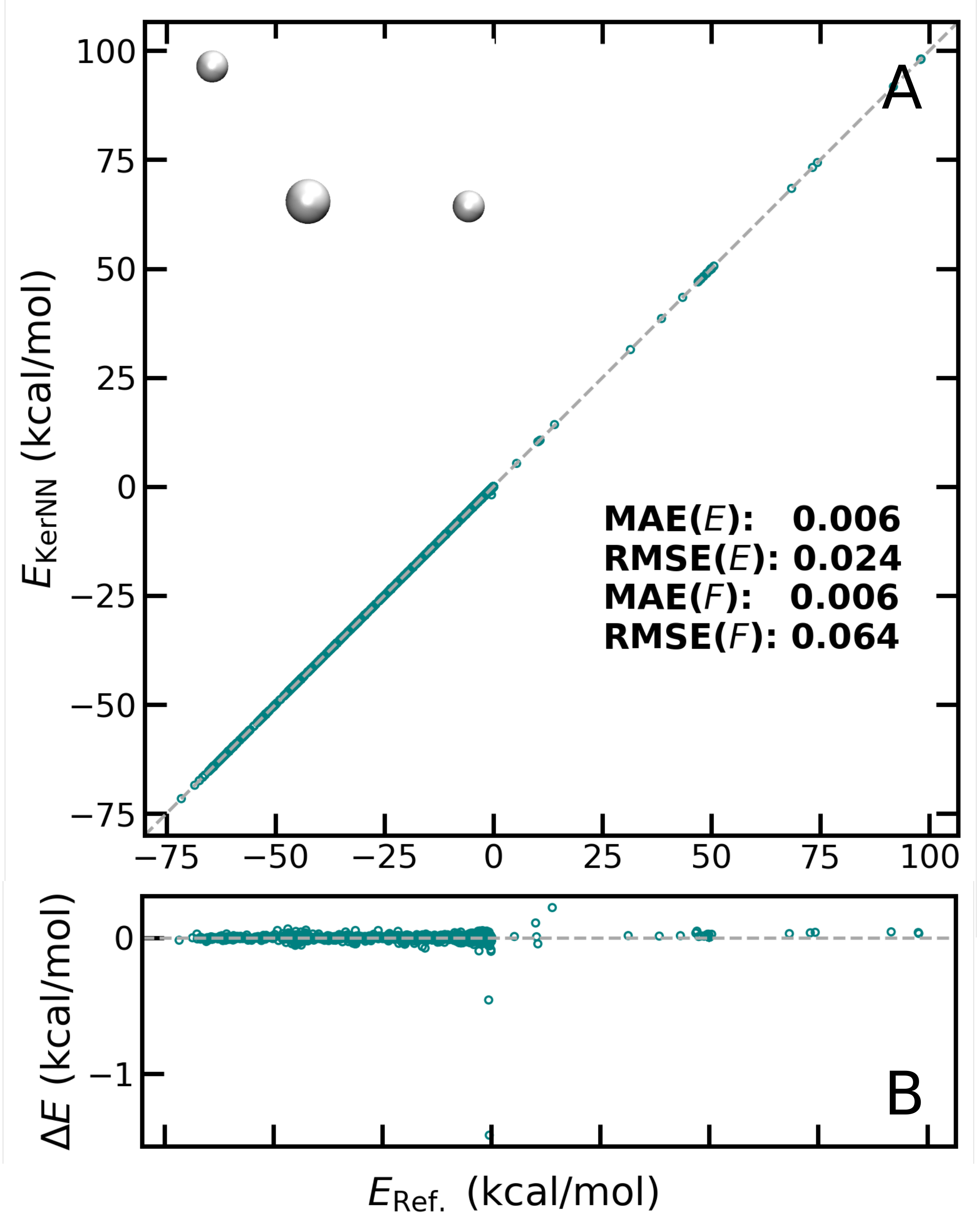}
\caption{Test set errors for the H + HeH$^+$ KerNN PES trained on
  UCCSD(T)/aug-cc-pV5Z level reference data\cite{ccsdt1994}. The test
  set contains 4784 randomly chosen structures. Most energies are
  predicted with errors well below 0.1~kcal/mol, while single outliers
  exist in the dissociative region. The test structures with largest
  $\Delta E = E_{\rm Ref.} - E_{\rm KerNN} > 1$ kcal/mol is shown in
  the main panel featuring $r(\rm{HeH}) = 4.34$ a$_0$. The zero of
  energy is the complete dissociation He + H + H$^+$.}
\label{fig:kernn_oos_errors_heh2+}
\end{figure}

\noindent
To characterize the quality of the KerNN representation, the
performance on the test set is shown in
Figure~\ref{fig:kernn_oos_errors_heh2+}. Across an energy range of 175
kcal/mol the mean absolute and root mean squared errors (MAE and RMSE)
on the energies are 0.006 and 0.024 kcal/mol, see
Figure~\ref{fig:kernn_oos_errors_heh2+}A. The correlation coefficient
between reference and KerNN-predicted energies is $R^2 =
1-10^{-6}$. For the forces, which are required to propagate the
classical equations of motion, the MAE and RMSE are 0.006 and 0.064
kcal/mol/\AA\/. For most structures in the test set the difference
between reference and model energies is well below 0.1 kcal/mol, see
Figure~\ref{fig:kernn_oos_errors_heh2+}B. For the one outlier with
$\Delta E > 1$ kcal/mol the structure is $\sim 75$ kcal/mol above the
global minimum and features long He--H and H$_{\rm A}$--H$_{\rm B}$
separations. Such open structures are prone to multi-reference effects
and UCCSD(T) as a single-reference method is potentially insufficient
to correctly describe their electronic structure.\cite{mm.kernn:2024}
It is also noted that for a three-electron system CCSDT and FCI are
equivalent which implies that UCCSD(T) is expected to be a good
approximation for both. This was explicitly confirmed by comparing
results from FCI/aug-cc-pV5Z calculations with the KerNN
representation of the UCCSD(T)/aug-cc-pV5Z data, see
Figure~\ref{sifig:pes-valid}.\\

\begin{figure}[H]
\centering
\includegraphics[width=1.0\textwidth]{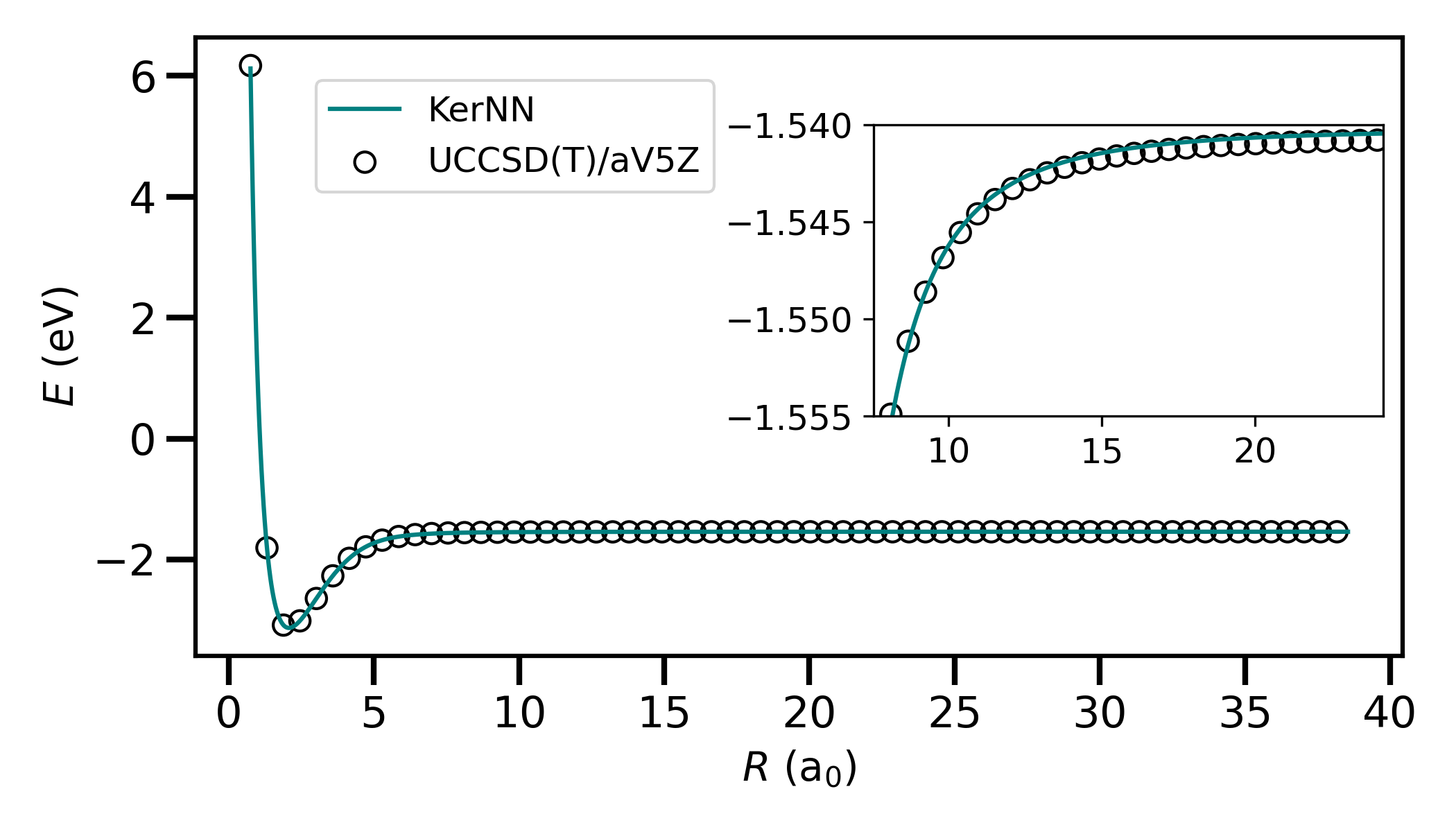}
\caption{One-dimensional potential energy scan along $R$ which is the
  distance between the two hydrogens, for a collinear geometry
  ($\theta = 0$) with $0.75 < R < 38.5$~a$_0$ and $r(\rm{HeH}^+) =
  1.94$~a$_0$. The energy range in the inset covers 0.34~kcal/mol
  (equivalent to 120 cm$^{-1}$) and all distances $R$ are within the range of the training data. The inset demonstrates that no artificial barrier is present.
  }
\label{fig:pes_scan_R}
\end{figure}

\noindent
A comparison with explicit electronic structure calculations is
provided for a one-dimensional scan along the collinear approach
HeH$^+$--H in Figure~\ref{fig:pes_scan_R}. This is particularly
relevant as earlier PESs\cite{ramachandran:2009,sahoo.refined.2024}
featured artificial barriers of different heights for H$_{\rm A}$
approaching the HeH$_{\rm B}^+$ ion in such a geometry. The inset of
Figure~\ref{fig:pes_scan_R} demonstrates that neither the
UCCSD(T)/aV5Z reference data nor its KerNN representation indicate any
barrier. Furthermore, the main view of Figure~\ref{fig:pes_scan_R}
underlines the excellent performance of KerNN compared with electronic
structure calculations. The differences between reference data and
KerNN representation remain well below 0.1 kcal/mol even for $R > 10$
a$_0$ (inset).\\

\begin{figure}[H]
    \centering \includegraphics[width=16.5cm]{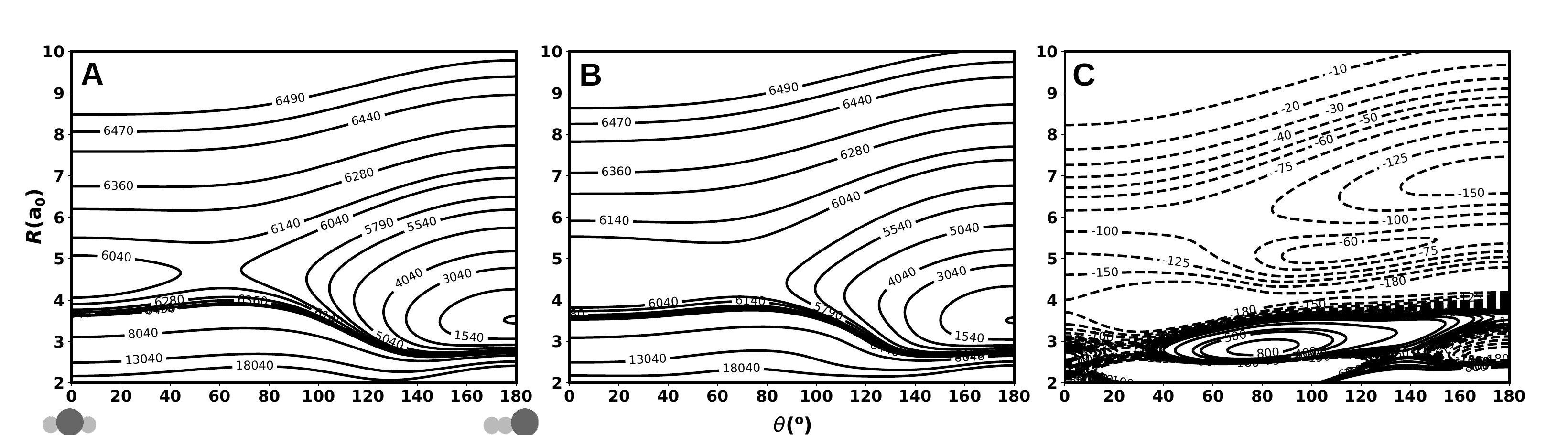}
    \caption{Two-dimensional PESs at fixed $r_{\rm HeH^+} = 1.46$
      a$_0$ as a function of $R$ (H--CoM(HeH$^+$) separation) and
      angle $\theta$ between the $r-$ and $R-$vectors. Panels A and B
      correspond to KerNN and cR-PES, while Panel C represents the
      difference $\Delta V(R,\theta) $ with solid and dashed lines
      representing positive and negative values, respectively. Both
      PESs uses the same set of contours, with minimum energy for this
      cut as the zero of energy at $R=3.52$ a$_0$ and $
      \theta=180^\circ$.}
    \label{fig:pes}
\end{figure}

\noindent
Figure \ref{fig:pes}A and B directly compare the KerNN and cR-PES for
the H + HeH$^+$ entrance channel at a fixed HeH$^+$ separation of
$r=1.46$ a$_0$. Both PESs use the same set of contours, with the zero
of energy (in cm$^{-1}$) defined at the global minimum with $R=3.52$
a$_0$ and $ \theta=180^\circ$. In Panel A, the minima at
$\theta=0^\circ$ and $180^\circ$ correspond to the [H-He-H]$^+$ and
[H-H-He]$^+$ conformations, respectively. The absence of a local
minimum at the linear structure in the cR-PES is due to the chosen
contours, which highlights significant differences in the shape of the
PESs in this region of configurational space. To better characterize
the differences, Panel C reports the PES difference $\Delta
V(R,\theta) = V_{\rm KerNN} (R,\theta) - V_{\rm cR-PES}(R,\theta)$,
where solid and dashed lines indicate positive and negative values,
respectively. Qualitatively, the cR-PES is more attractive in the long
range [$R=3 - 10$ a$_0$] and more repulsive in the short range
compared to the KerNN PES.\\

\begin{figure}[H]
    \centering \includegraphics[width=14.5cm]{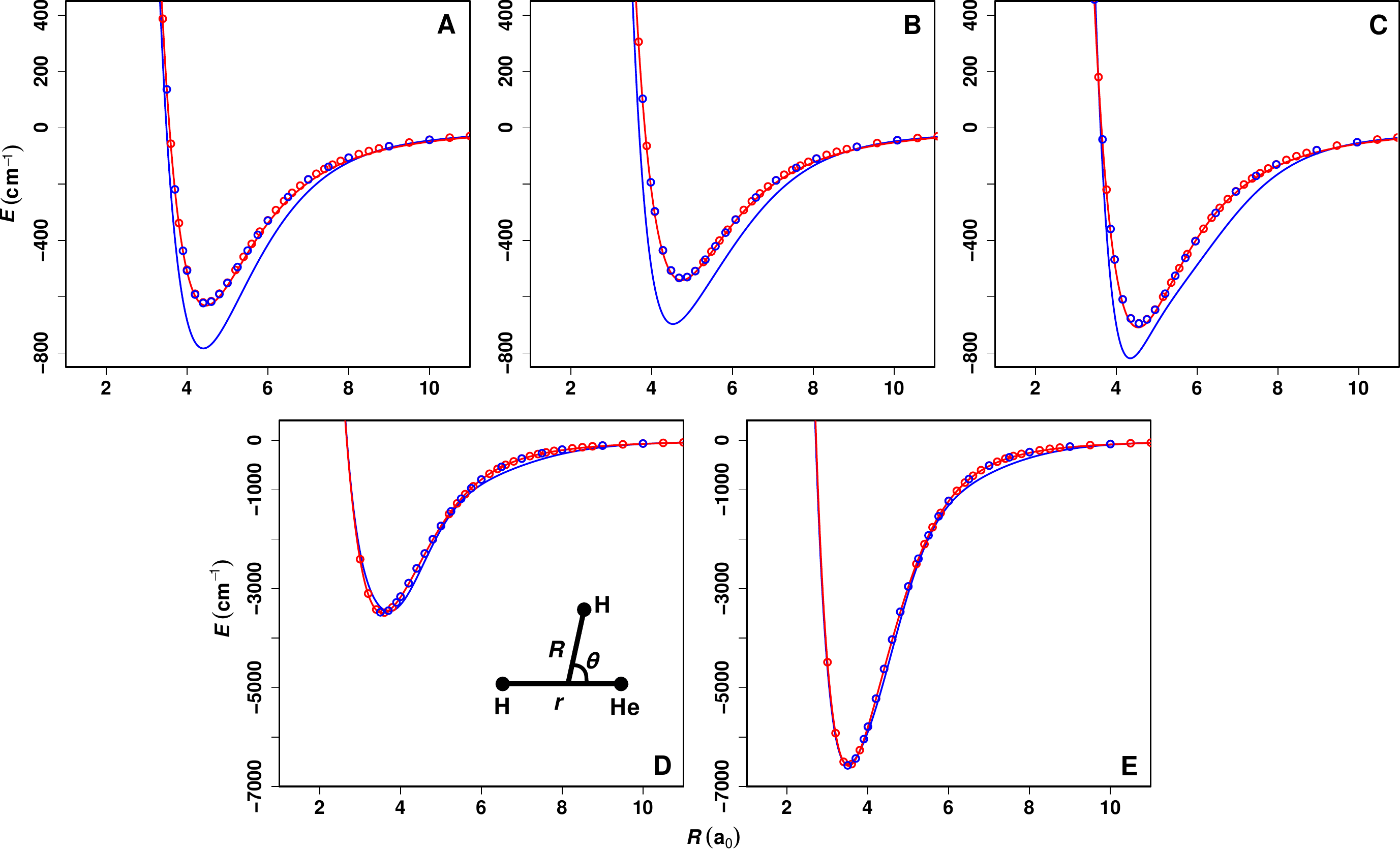}
    \caption{One-dimensional off-grid cuts for [H--He--H]$^+$ at fixed
      $\theta$ and for $r({\rm He-H}) = 1.46$~a$_0$ from KerNN PES
      (red) and the adjusted surface cR-PES (blue). A:
      $\theta=0^\circ$, B: $\theta=45^\circ$, C: $\theta=90^\circ$, D:
      $\theta=135^\circ$, E: $\theta=180^\circ$. Here $\theta =
      0^\circ$ corresponds to the [H-He-H]$^+$ conformation and
      $\theta = 180^\circ$ corresponds to [He-H-H]$^+$. The bottom of
      HeH$^+$ diatom well is set as zero of energy. The open circles
      represent the UCCSD(T)/aug-cc-pV5Z (red) and MRCI/aug-cc-pV6Z
      (blue) \textit{ab initio} energies.}
    \label{fig:1dcut}
\end{figure}

\noindent
Finally, the KerNN and cR-PES are compared with reference calculations
at the UCCSD(T)/aug-cc-pV5Z and MRCI/aug-cc-pV6Z levels of theories;
see Figure \ref{fig:1dcut}. The 1-dimensional cuts give $V(R)$ for
HeH$^+$ separation $r = 1.46$ a$_0$ and for different angles of
approach of the second hydrogen atom. The KerNN representation (red
solid lines) is in excellent agreement with the explicit electronic
structure calculations (red open circles). On the other hand, for the
cR-PES energy function (blue line) the agreement is encouraging for
angles $\theta = 135^\circ$ and $\theta = 180^\circ$. For $\theta \leq
90^\circ$ (panels D and E), however, the cR-PES fails to capture the
correct $R-$dependence, both in terms of depth and shape (see
e.g. panel C). The failure of the cR-PES, particularly in the region
$\theta \leq 90^\circ$, can be attributed to deficiencies in the
three-body term which was that from the RFCI8
PES.\cite{ramachandran:2009} However, it is important to note that the
cR-PES features no barrier in the H + HeH$^+$ entrance channel by
incorporating accurate two-body long-range interaction terms from the
literature.\cite{sahoo.refined.2024} \\

\subsection{Reactive Classical and Quantum Simulations}
Reactive QCT simulations were carried out with the KerNN and cR-PES
surfaces. To determine suitable conditions for running the QCT
simulations, first the opacity function $P(b)$ for the ${\rm HeH}^+
(v=0,j=0) + {\rm H} \rightarrow {\rm H}_2^+ + {\rm He}$ reaction was
determined at different temperatures, see Figure
\ref{sifig:opacity}. The opacity function characterizes the total
reaction probability as a function of the impact parameter $b$. For
$T=100$ K the two PESs KerNN (red) and cR-PES (blue) yield comparable
$P(b)$ with $b_{\rm max} \sim 25$ a$_0$. However, decreasing the
temperature to $T = 10$ K markedly extends the required range of
impact parameters to be sampled to $b > 30$ a$_0$ for both PESs.\\

\begin{figure}[H]
    \centering
    \includegraphics[width=15.5cm]{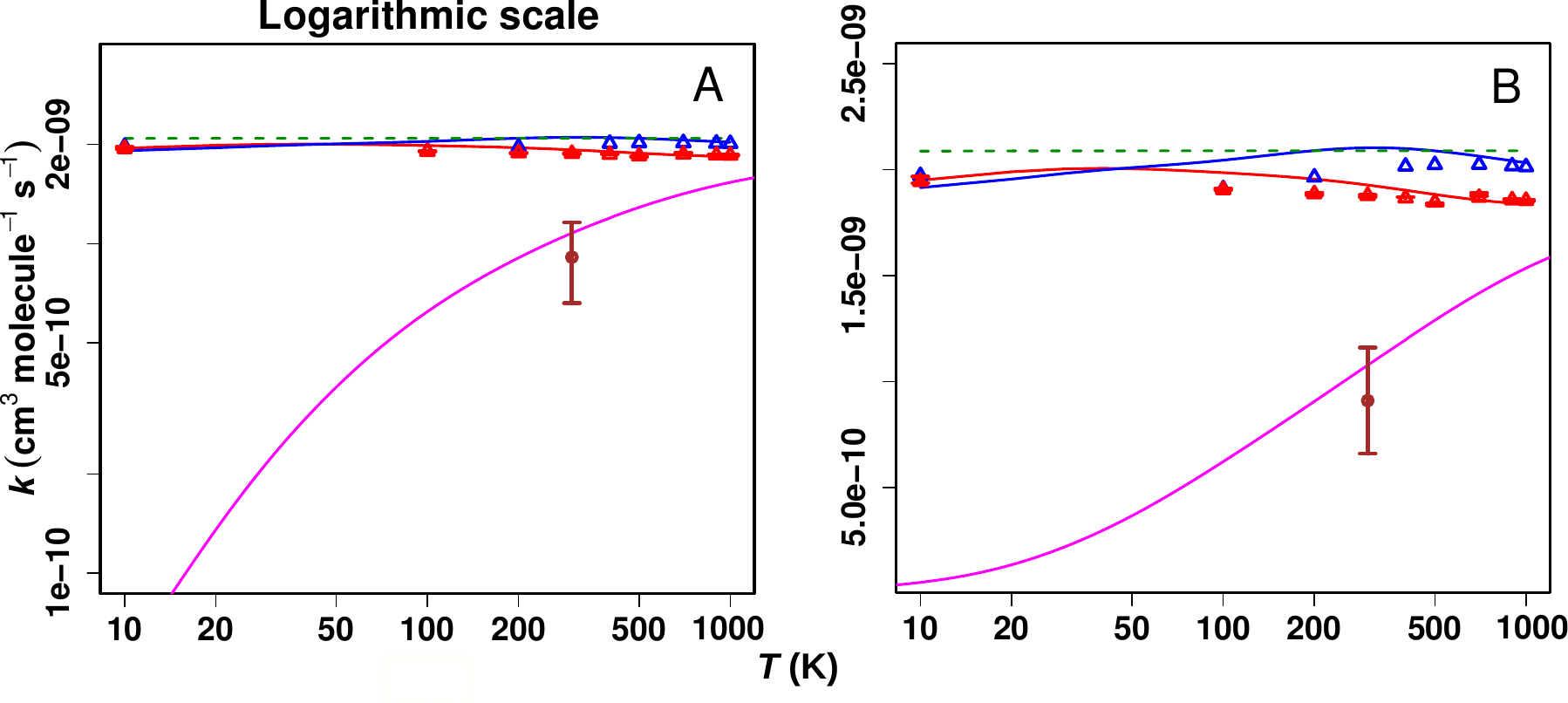}
    \caption{Initial state-selected $T-$dependent rates $k(T)$ for the
      ${\rm H} + {\rm HeH}^+(v=0,j=0) \rightarrow {\rm He} + {\rm H}_2^+ $ reaction. Panels A and B report $k(T)$ on logarithmic and
      linear scales, respectively. Solid lines correspond to the TIQM
      simulations using the KerNN (red), cR-PES (blue) and the RMRCI6
      (magenta) PESs.\cite{ramachandran:2009} The QCT results are the
      red and blue triangles, respectively. The brown circle with the
      error bar is the ICR-measured rate\cite{karpas:1979} $k_{\rm
        ICR} = (9.1 \pm 2.5) \times 10^{-10}$ cm$^3$/molecule/s and
      the dashed green line is the classical Langevin result $k_{\rm
        L} = 2.1\times 10^{-9}$ cm$^3$/molecule/s. Differences in the
      rate coefficients between QCT and TIQM simulations are larger
      for the cR-PES compared with KerNN PES. Error bars for the
      QCT-results from bootstrapping are smaller than the symbol
      size.}
    \label{fig:rates}
\end{figure}

\noindent
Temperature-dependent rates as experimental observables provide a
meaningful benchmark for the PESs and dynamics simulations using
them. The computed $k(T)$ from QCT simulations using the KerNN and
cR-PES surfaces are reported in Figure \ref{fig:rates}. The QCT
simulations were carried out for the H + HeH$^+ (v=0, j=0)$
$\longrightarrow$ He + H$_2^+$ (all $v,j$) reaction using $b_{\rm max}
= 30$ a$_0$ and $r_{\rm 0} = 32$~a$_0$. With both PESs (red and blue
symbols) $k_{v,j}(T)$ is flat and varies little with
temperature. Rates from using the KerNN PES (red) are somewhat smaller
than those from using the cR-PES (blue) which can be best seen in
Figure \ref{fig:rates}B. However, both sets of rates are consistent
with the Langevin rate (green dashed line) which often serves as a
proxy for a barrierless reaction with given charge-induced dipole
interaction at long range.\cite{langevin:1905formule,gio58:294}\\

\noindent
Quantum simulations were also carried out using both PESs. For the
KerNN (red) and cR-PES (blue) the rates from TIQM (solid lines) and
QCT (triangles) simulations agree very favourably, even at the lowest
temperatures considered ($T = 10$ K). For the KerNN PES the agreement
between QCT and TIQM simulations is even somewhat better. This is also
consistent with previous work on the O + CO reaction which also found
that thermal rates from quantum and classical dynamics simulations are
consistent with one another.\cite{MM.co2:2021} As an additional
comparison, results from earlier TIQM simulations\cite{fazio:2014}
using the RMRCI6 PES are reported as the solid magenta line in Figure
\ref{fig:rates}. Clearly, the spurious barrier in this PES suppresses
$k(T)$ at low temperatures when compared with the present
simulations.\\

\noindent
Experimentally, the only reported measurement for the H + HeH$^+$
 reaction rate is based on an ion cyclotron resonance study
(ICR)\cite{karpas:1979} using a microwave discharge for generating the
ions. The reported value (at unspecified temperature - assumed to be
300 K in the subsequent literature) to form He + H$_2^+$ is $k_{\rm
  ICR} = (9.1 \pm 2.5) \times 10^{-10}$ cm$^3$/s with probability 1
(brown symbol including error bars in Figure
\ref{fig:rates}). However, it should be noted that the reported rate
is not an absolute rate but was measured relative to a rate for the
reaction HeH$^+$ + H$_2$ to form H$_3^+$ + He. Furthermore, serious
doubts on the validity of these ICR measurements had been cast from
work on the CO$_2^+$ + H$_2$ $\rightarrow$ HCO$_2^+$ + H reaction and
H-isotopic analogs.\cite{gerlich:2009} Comparison with measurements in
a 22-pole trap indicated that the ICR measurements underestimate rates
by at least a factor of 5 for this
reaction.\cite{gerlich:2009,karpas:1979} Hence, $k_{\rm ICR} = (9.1
\pm 2.5 ) \times 10^{-10}$ cm$^3$/molecule/s for the H + HeH$^+$
reaction is a lower limit. Assuming the same factor of 5 as for the
CO$_2^+$ + H$_2$ $\rightarrow$ HCO$_2^+$ + H reaction yields $k_{\rm
  expt}^{\rm estimate} \approx (4.55 \pm 2.5) \times 10^{-9}$
cm$^3$/molecule/s for the H + HeH$^+$ reaction. Including error bars,
this is consistent with the Langevin-rate $k_{\rm L} = 2.1\times
10^{-9}$ cm$^3$/molecule/s and the QCT and TIQM simulations based on
KerNN and cR-PES surfaces.\\

\begin{figure}
    \centering
    \includegraphics[width=10.5cm]{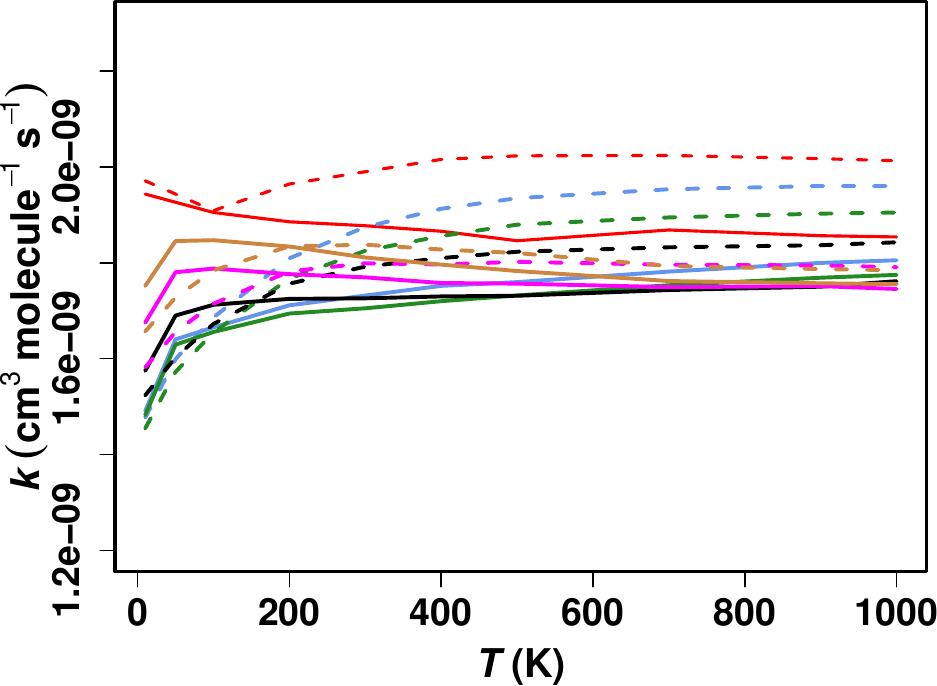}
    \caption{Initial state-selected $T-$dependent rates $k_{v,j}(T)$
      from QCT simulations for the ${\rm HeH}^+(v=0,j) + {\rm H}
      \rightarrow {\rm H}_2^+ + {\rm He}$ reaction with $b_{\rm max}=
      30$ and $r_0=32$ a$_0$. Color code for the initial conditions of
      HeH$^+$: $j=0$ (red), $j=1$ (blue), $j=2$ (green), $j=3$
      (black), $j=4$ (magenta), and $j=5$ (brown). Solid and dashed
      lines correspond to simulations using the KerNN and cR-PES
      surfaces, respectively. For results from the TIQM simulations,
      see Figure \ref{sifig:tiqm-rates-j0j5}.}
    \label{fig:rates-j0to5}
\end{figure}

\noindent
Next, additional initial state-selected rate coefficients were also
computed from QCT simulations using the KerNN and cR-PES surfaces, see
Figure \ref{fig:rates-j0to5}. The state-selected rates for $j=4$ and
$5$, computed using the KerNN PES, initially increase with temperature
and then gradually decay towards a limiting value of $\sim 1.8 \times
10^{-9}$ cm$^3$ molecule$^{-1}$ s$^{-1}$ at $T = 1000$ K. In contrast,
the rates for $j=0$ exhibit a negative $T-$dependence, while those for
for $j=1,2$ and $3$ a positive $T-$dependence was
observed. Simulations using the cR-PES exhibit a positive
$T-$dependence for all initial $j-$values except for $j=0$. Also, the
limiting values using cR-PES differ from one another and from those
obtained with the KerNN PES. Qualitatively, initial state-selected
$k(T)$ from TIQM simulations, see Figure \ref{sifig:tiqm-rates-j0j5},
exhibit the same $T-$dependence and limiting values in the
high-temperature limit. However, for the KerNN PES there is a local
maximum in $k(T)$ for $T < 100$ K which is not found from simulations
using the cR-PES. Again, except for initial $j=0$, $k(T)$ from
simulations using the KerNN PES converge to a common value in the
high-$T$ limit which is again not the case for the cR-PES. Also, the
range of $k(T)$ covered is similar at higher temperatures but the
lower limit differs for lower temperatures. Although measurements are
available for integral cross sections,\cite{vroom:1973} these were not
calculated here because they did not distinguish between different
PESs at experimentally relevant energies, see Figure 6 (left) of
Ref.\citenum{sahoo.refined.2024}, where the integral cross sections
for collision energies $> 0.2$ eV from TIQM simulations using the R-
and cR-PESs are virtually indistinguishable.\\

\begin{figure}[H]
    \centering
    \includegraphics[width=12.5cm]{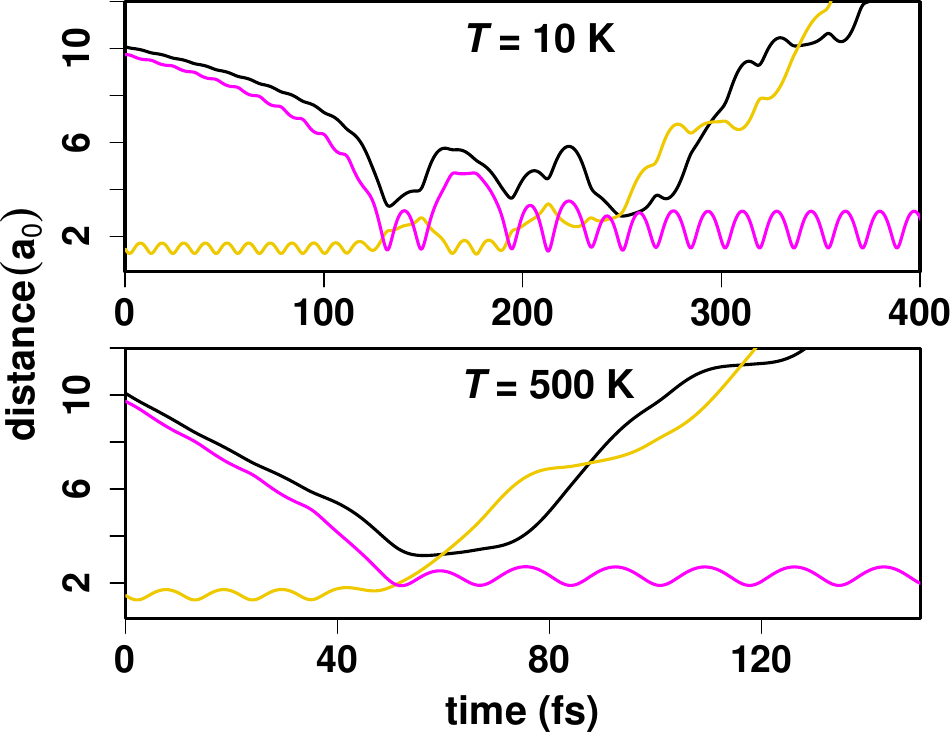}
    \caption{Time evolution of the internuclear distances from QCT
      simulation on the KerNN PES of the diatoms, He--H (black and
      gold) and H--H (magenta), as a function of time for the H + HeH$^+
      (v=0, j=0)$ $\longrightarrow$ He + H$_2^+$ reaction at $T=$
      10 and 500 K.}
    \label{fig:qct_traj}
\end{figure}

\noindent
To obtain a direct impression of how the reaction proceeds at an
atomistic level it is of interest to consider time series of
internuclear separations. Two representative trajectories from
simulations using the KerNN PES at $T = 10$ K and $T = 500$ K are
shown in Figure \ref{fig:qct_traj}. The trajectory at low temperatures
forms a [HHeH]$^+$ collision complex with a lifetime of $\sim 150$ fs
, corresponding to a few $(< 10)$ vibrational periods. Conversely, the
high-temperature trajectory features an H-abstraction reaction within
a single collision with contact times of $\sim 30$ fs.\\

\begin{figure}
    \centering
    \includegraphics[width=0.95\linewidth]{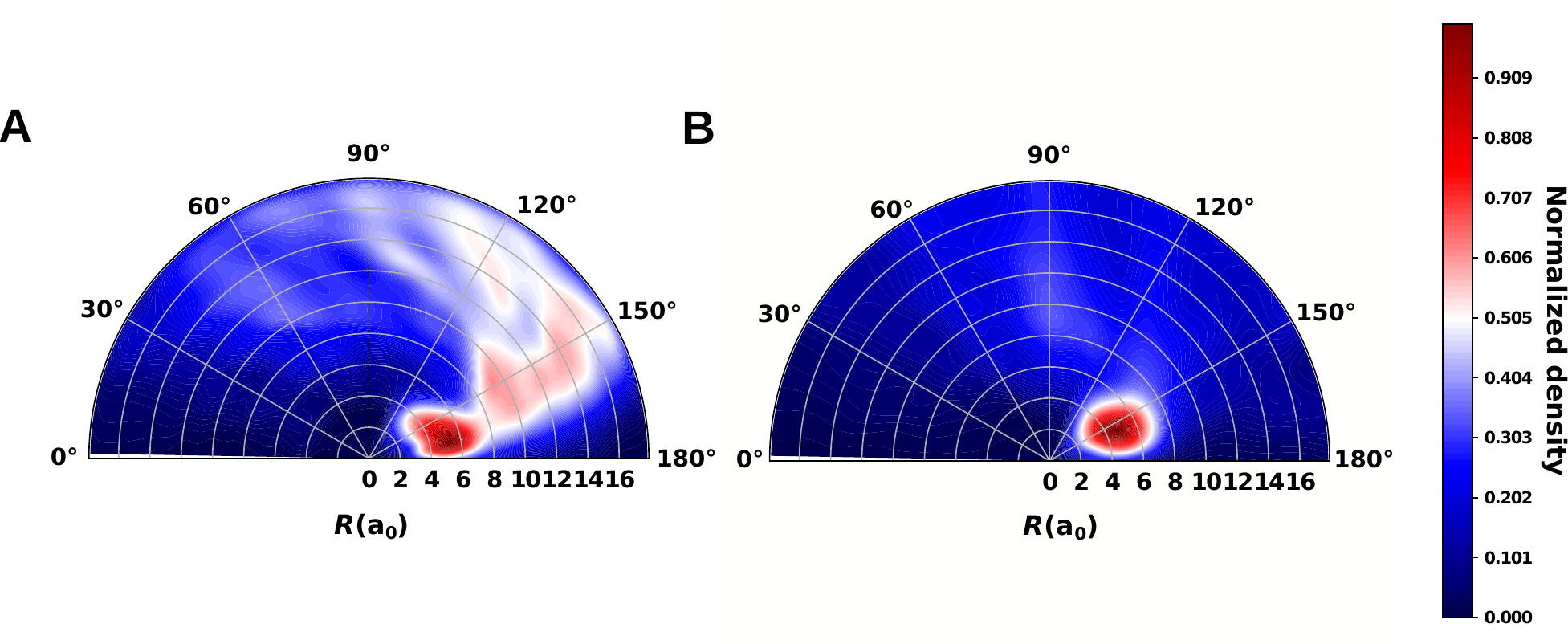}
    \caption{Geometrical sampling of the incoming H-atom around the
      center of mass of He-H$^+$ before and during complex formation
      (i.e before H-transfer) for 5000 trajectories using the KerNN
      (panel A) and cR-PES (panel B) for $T= 100$ K and with $r_0=32$
      a$_0$. The sampling frequency of data points is $\Delta t =0.75$
      fs. The [H-He-H]$^+$ and [He-H-H]$^+$ structures correspond to
      $\theta = 0^\circ$ and $\theta = 180^\circ$, respectively.}
\label{fig:full_geo}
\end{figure}

\noindent
A more comprehensive picture of the collision process can be obtained
from analyzing a distribution of trajectories. For this, 5000 reactive
H$_{\rm A}$ + HeH$^+_{\rm B} (v=0, j=0)$ + $\longrightarrow$ He +
H$_2^+$ trajectories from simulations using the KerNN and cR-PES
surfaces and run at 100 K were selected. All $(R,\theta)$ pairs were
recorded for $r_{\rm He-H_{\rm B}} \leq 6$
a$_0$, i.e., before H-transfer and during complex formation, see
Figure \ref{fig:qct_traj}A.  As the geometry criterion was applied to
the value of $r$, arbitrary values for $R$ and $\theta$ are
possible. The distribution $P(R,\theta)$ was then projected onto the
$(R,\theta)-$plane, see Figure \ref{fig:full_geo}, with $\theta =
0^\circ$ and $\theta = 180^\circ$ corresponding the [H-He-H]$^+$ and
[He-H-H]$^+$ conformations, respectively. In Figure
\ref{fig:qct_traj}A, the diffusive motion of the H-atom governed by
long-range interaction can be clearly seen. As the incoming H-atom
approaches to within $R \sim 10$ a$_0$ channeling towards the HeH$^+$
collision partner occurs which eventually leads to formation of the
[He-H-H]$^+$ collision complex and breakup into He and H$_2^+$. In
contrast, this diffusive channeling is not observed in panel B, which
presents results from simulations using the cR-PES. This is likely due
difference in the long-range part of the cR-PES compared with the
KerNN PES in particular for the [H-He-H]$^+$ geometry, see Figures
\ref{fig:pes} and \ref{fig:1dcut}. The diffusive nature of the
incoming hydrogen motion is due to the comparatively less attractive
nature of the long-range interactions in the KerNN PES. A more
comprehensive rendering of the dynamics during complex formation at
different temperatures is shown in Figure \ref{sifig:complex-geo}.\\

\noindent
It is also of interest to consider the lifetimes of the [HeH$_2^+$]
complex for trajectories showing H-transfer, i.e H + HeH$^+$
$\longrightarrow$ He + H$_2^+$. Here, the lifetime $\tau$ of the
collision complex was defined as the time elapsed from the first
instance when each interatomic separation is smaller than $6$ a$_0$
until the moment when any one of the separations is larger than $6$
a$_0$. The normalized lifetime distributions $P(\tau)$ at different
temperatures are shown in Figure \ref{sifig:lifetime}. Trajectories at
higher temperatures have longer lifetimes due to formation of tightly
bound complexes, roaming of He around H$_2^+$ before complete
dissociation and/or undergoing the reaction sequence before complete
fragmentation: ${\rm H_A} + {\rm HeH_B^+} \rightarrow {\rm He} \cdots {\rm H}_2^+
\rightarrow {\rm H_B} \cdots {\rm HeH_A^+} \rightarrow
            {\rm He} + {\rm H}_2^+$.  \\

\subsection{Final State Distributions from both PESs}
\begin{figure}[H]
    \centering \includegraphics[width=16.5cm]{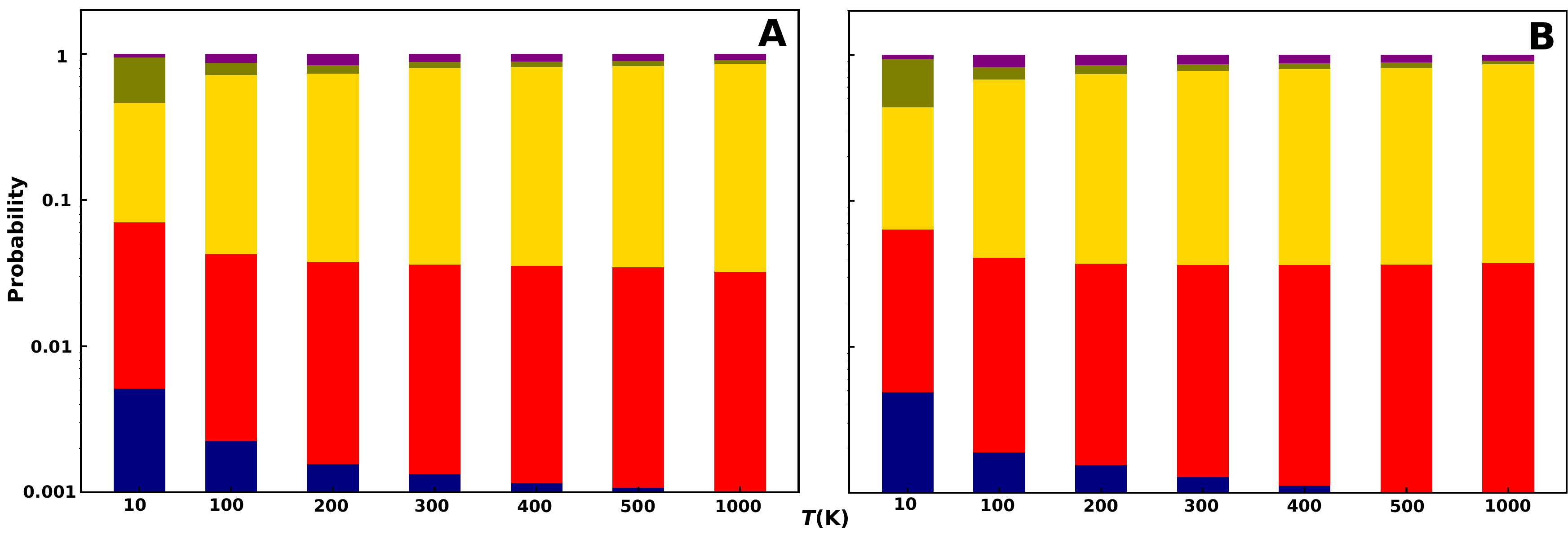}
    \caption{Relative fraction of the different possible reaction
      channels on a logarithmic $y-$scale (to also show
      low-probability channels) using KerNN (left) and cR-PES (right)
      depending on temperature. The relevant processes and scenarios
      considered include a) H-transfer (green): H$_{\rm A}$ + HeH$_{\rm B}^+  \longrightarrow$ He +
H$_2^+$; b) Atom exchange
      (blue): H$_{\rm A}$ + HeH$_{\rm B}^+  
\longrightarrow$  HeH$_{\rm A}^+ $ + H$_{\rm B}$; c) Inelastic Collisions (red): H$_{\rm A}$ + HeH$_{\rm B}^+ (v=0,j=0) \longrightarrow$ H$_{\rm A}$ + HeH$_{\rm B}^+ (v \ne 0, j \ne 0)$; d) Elastic collisions
      (purple): H$_{\rm A}$ + HeH$_{\rm B}^+ (v=0,j=0)
\longrightarrow$ H$_{\rm A}$ + HeH$_{\rm B}^+ (v=0,j=0)$; e)
      Flyby (yellow). See Figure \ref{sifig:prob-kernn-linear} for
      linear scaling along the $y-$axis.}
    \label{fig:prob-kernn-log}
\end{figure}

The final state distributions $P(v')$ and $P(j')$ of the H$_2^+
(v',j')$ fragment also contain valuable information. Because the
outcome of the collision process includes multiple channels, first an
overview of all possible final states are given. These include a)
H-transfer (green): H$_{\rm A}$ + HeH$_{\rm B}^+  \longrightarrow$ He +
H$_2^+$, b) Atom exchange (blue):  H$_{\rm A}$ + HeH$_{\rm B}^+  
\longrightarrow$  HeH$_{\rm A}^+ $ + H$_{\rm B},$ c) Inelastic
Collisions (red): H$_{\rm A}$ + HeH$_{\rm B}^+ (v=0,j=0)
\longrightarrow$ H$_{\rm A}$ + HeH$_{\rm B}^+ (v \ne 0, j \ne 0)$, d)
Elastic collisions (purple): H$_{\rm A}$ + HeH$_{\rm B}^+ (v=0,j=0)
\longrightarrow$ H$_{\rm A}$ + HeH$_{\rm B}^+ (v=0,j=0)$, and e)
Flyby (yellow).  Starting from the H$_{\rm A}$ + HeH$_{\rm B}^+$
reactant, Figure \ref{fig:prob-kernn-log} shows the relative
probabilities for a) formation of H$_2^+$ (green, [5 to 45] \%), b)
H-atom exchange (blue, $ < 1$ \%), c) inelastic collisions (red, $< 1$
\%), d) elastic collisions (purple, typically [5 to 20] \%), and e)
flyby (yellow, typically [40 to 85] \%). A logarithmic scale was used
to make low-probability processes visible within the same chart. For a
linear $y-$scale, see Figure \ref{sifig:prob-kernn-linear}.\\

\noindent
For temperatures below 100 K the probability for H$_2^+$ formation is
close to 50 \% whereas this fraction decreases rapidly for $T \geq
100$ K. At higher temperatures most trajectories correspond to
``flyby". The fraction of inelastic collisions is below 10 \%
throughout and decreases with increasing temperature, whereas for
elastic collisions, the probability initially increases up to 200 K
and then declines with increasing temperature. Overall, the two PESs
yield comparable probabilities for the five channels which is reassuring
given that the way in which the PESs were conceived is rather
different.\\

\begin{figure}[H]
    \centering
    \includegraphics[width=1.0\linewidth]{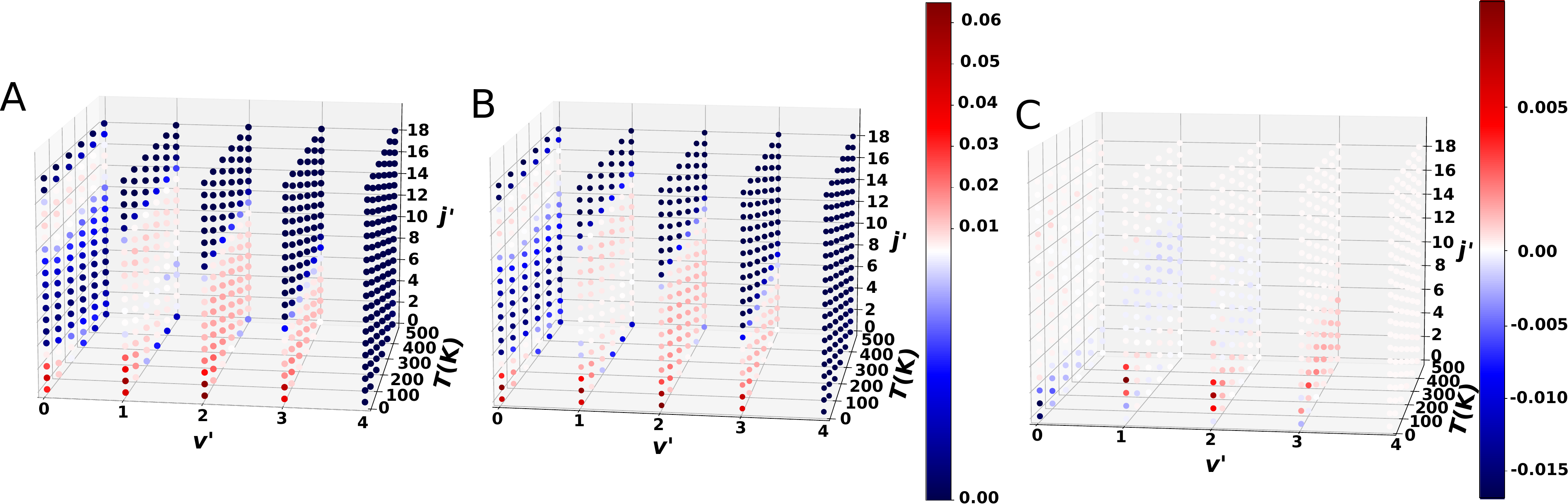}
    \caption{Probability of final ${\rm H}_2^+(v',j')$ states starting
      from initial H + HeH$^+$($v=0, j=0$) to form He + H$_2^+$ for $T
      = 10$ to 500 K, using the KerNN (panel A) and the cR-PES (panel
      B). The differences $\Delta P(v',j') = P_{\rm KerNN}(v',j') -
      P_{\rm cR-PES}(v',j')$ are reported in panel C.}
    \label{fig:pvj_T}
\end{figure}

\noindent
Next, using both reactive PESs the final $(v',j')$ states for the
H$_2^+$ product were analyzed by way of final state distributions
$P(v')$ and $P(j')$. This is of interest to delineate the state-space
of the reaction product H$_2^+(v',j')$ because it is known that
different internal states of molecules can modulate their downstream
reactivity. For this, QCT simulations were run with different
collision energies drawn from a Boltzmann distribution at a given
temperature. The final state distributions are reported in Figure
\ref{fig:pvj_T}A and B and show that the two PESs yield different
probability distributions and the differences are reported in Panel
C. In particular, the population of $P(v' = 0)$ is less likely for
KerNN compared with simulations using the cR-PES. This can be
rationalized by referring to Figure \ref{fig:1dcut} which indicates
that on the KerNN PES -- which is consistent with the UCCSD(T)
reference data -- more energy is released (and therefore available)
upon formation of He--H$_2^+$ for an approach between $\theta = 0$ and
$90^\circ$. Another noteworthy observation is that the differences in
the probabilities remain small $\in [-0.004, 0.003]$ in panel C,
except for lower $v', j'$ and $T$, for which deviations increase due
to sampling parts of the PES where differences in the KerNN and cR-PES
are more pronounced. On the other hand the differences diminish as $T$
increases which is consistent with the findings for the geometrical
sampling in Figure \ref{sifig:complex-geo}. For a conventional
1-dimensional $P(v')$ from QCT and TIQM simulations see Figures
\ref{sifig:1d-pv} and \ref{sifig:pv-quantum}.\\

\begin{figure}
    \centering
    \includegraphics[width=11.5cm]{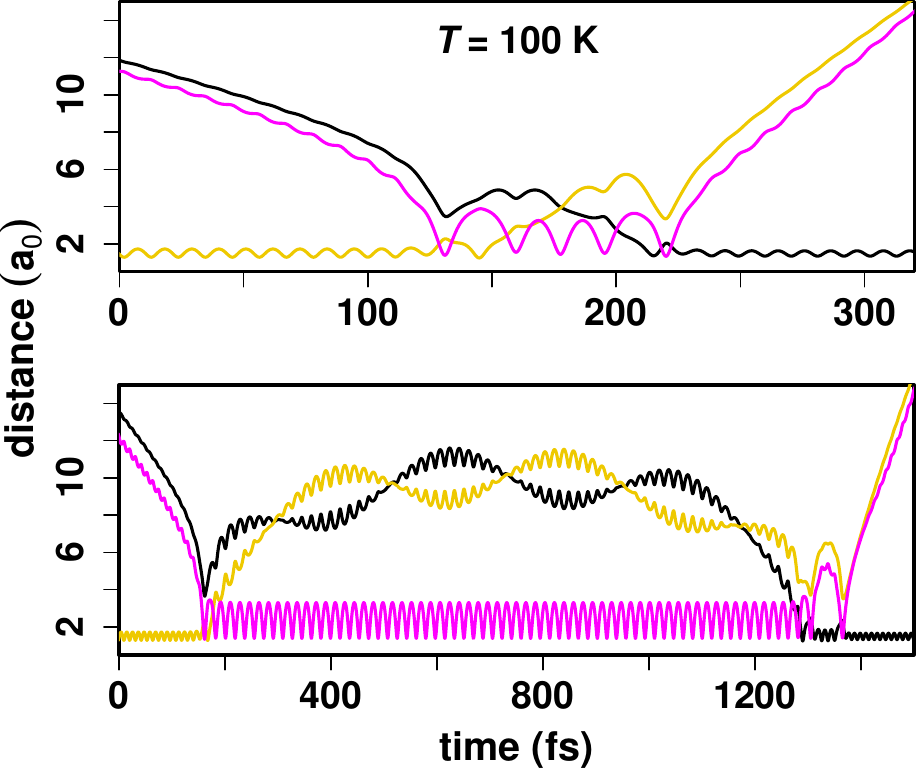}
    \caption{Time evolution of the internuclear separations for H-atom
      exchange reactions from QCT simulation on the KerNN PES of the
      diatoms, He--H (black and gold) and H--H (magenta), as a
      function of time for $T= 100$ K. The lower panel shows a
      ``roaming trajectory''.}
    \label{fig:qct_traj2}
\end{figure}

\noindent
The QCT simulations find that different final states can be reached
for the H$_{\rm A}$ + HeH$_{\rm B}^+$ reaction. Up until this point
the final channel  H$_{\rm A}$ + HeH$_{\rm B}^+  \longrightarrow$ He +
H$_2^+$ was considered specifically (green bar in Figure
\ref{fig:prob-kernn-log}). The contribution of this channel decreases
with increasing temperature. A process that increases in proportion
with increasing temperature is the inelastic channel H$_{\rm A}$ + HeH$_{\rm B}^+ (v=0,j=0) \longrightarrow$ H$_{\rm A}$ + HeH$_{\rm B}^+ (v \ne 0, j \ne 0)$. Here the reactant and product states are
identical in a chemical sense but the product HeH$_{\rm B}^+$ is
rovibrationally excited. Also, the probability for elastic collisions
increases from 5 to 20 \% up to 200 K before decreasing to 9 \% at
1000 K. In such trajectories, a change in translational energy of the
approaching atom before and after the collision is observed. The most
likely channel at almost all temperatures is the ``flyby" channel (yellow in
Figure~\ref{fig:prob-kernn-log}). Finally, it is possible that H-atom
exchange occurs (blue distribution): H$_{\rm A}$ + HeH$_{\rm B}^+  
\longrightarrow$  HeH$_{\rm A}^+ $ + H$_{\rm B}$. This channel also
decreases in proportion as temperature increases. Even at the lowest
temperature considered (10 K) its contribution is less than 1
\%. Nevertheless, it is remarkable that H-atom exchange is found at
all, given that the He--H$_2^+$ complex is stabilized by more than 1
eV relative to the H + HeH$^+$ entrance channel and dissociation to He +
H$_2^+$ only requires 0.34 eV. An explicit trajectory for H-atom
exchange is shown in Figure \ref{fig:qct_traj2}.\\

\subsection{TKER Spectra for H + HeH$^+$  $\longrightarrow$ He + H$_2^+$}
\begin{figure}
    \centering
    \includegraphics[width=0.8\linewidth]{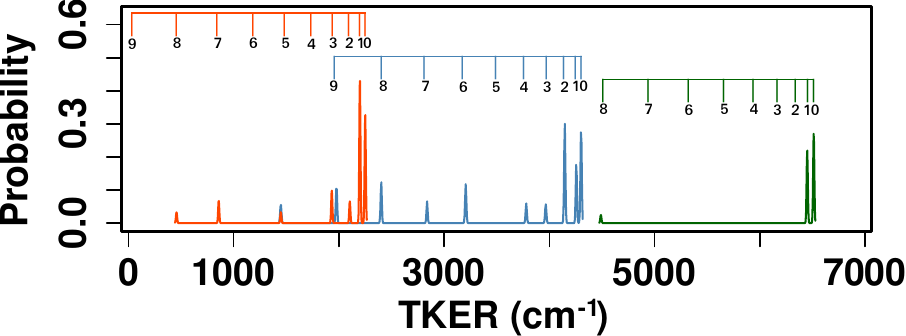}
    \caption{Total kinetic energy release (TKER) for He -- H$_2^+$
      product after H + HeH$^+$ ($v=0, j=0$) collision at $T=10$
      K. Simulations are carried out using the KerNN PES. Color code:
      Green: $v'=0$, blue: $v'=1$, red: $v'=2$.}
    \label{fig:tker}
\end{figure}

Finally, the H$_2^+$ product kinetic energies were analyzed, see
Figure \ref{fig:tker}. For this, only QCT simulations using the KerNN
PES were carried out as it is the globally valid description of the
[HHHe]$^+$ reactive collision system. Such translational kinetic
energy spectra have recently been reported for the breakup of the
He--H$_2^+$ complex which had been formed {\it in situ} through
Penning ionization.\cite{MM.heh2:2023} It has been found that the
Feshbach resonances underlying this process are particularly sensitive
to the long-range part of the PES.\cite{MM.morphing:2024} Hence, the
total kinetic energy release for the product of the H + HeH$^+ (v=0, j=0)$ $\rightarrow$ He + H$_2^+$ reaction was investigated at 10 K using
the KerNN PES (see SI for the method).\\

\noindent
The resulting kinetic energy spectra exhibit distinct peaks, each
corresponding to different final vibrational ($v' \in [0,2]$) and
rotational ($j' \in [0,9]$) states of the H$_2^+$ product. The
$j'-$states were assigned by solving the 1D time-independent
Schr\"{o}dinger equation using the LEVEL\cite{level:2017} program. In
particular, lower $v'$ and $j'$ states leads to a higher kinetic
energy released in the H$_2^+$ product.\\

\section{Discussion and Conclusions}
The present work introduced a precise KerNN-based, reactive PES
from UCCSD(T)/aug-cc-pV5Z reference data and used it in rigorous
time-independent quantum scattering and QCT studies. Because UCCSD(T)
is an approximation to CCSDT - which is equivalent to FCI for a three
electron system - the present {\it reactive} PES is among the most
accurate ones available so far. This aspect was explicitly and
quantitatively tested, see Figure \ref{sifig:pes-valid}. Also, it was
shown that for the H--HHe$^+$ approach the KerNN and cR-PES surfaces
agree quite closely, whereas for the higher-energy H--HeH$^+$ geometry
they do not.\\

\noindent
Given the caveats\cite{gerlich:2009} on the validity of the thermal
rate from the ICR-measurements\cite{karpas:1979} and the fact that QCT
and quantum simulations on two different PESs yield consistent results
(see Figure \ref{fig:rates}), the reported value $k_{\rm ICR} = (9.1
\pm 2.5) \times 10^{-10}$ cm$^3$/molecule/s is likely to be too low by
a factor of $\sim 5$. Also, the computed rates\cite{esposito:2015}
which ``reproduce" the ICR-measured rate and have been used in
modeling astrophysical reaction networks\cite{faure:2024} should
probably be replaced by rates $k(T)$ from QCT or TIQM simulations on
the KerNN PES which are consistent with the Langevin rate $k_{\rm L} =
2.1\times 10^{-9}$ cm$^3$/molecule/s. This is akin to recent findings
for the N+NO reaction for which a combination of experiment and
simulations led to proposing updated low-temperature reaction
rates\cite{MM.n2o:2023} which were subsequently adopted in astrophysical reaction networks.\cite{faure:pcomm,faure:2022,faure:2013} \\

\noindent
It is worthwhile to mention that following established
routes\cite{MM.no2:2020,MM.n2o:2020,MM.co2:2021} for constructing
reactive PESs for triatomics was not successful for the present
case. In other words, mixing the two asymptotic PESs for the
He--H$_2^+$ and H--HeH$^+$ PESs using exponential weighting was not a
viable approach to obtain a valid, accurate and global reactive PES as
had been possible for example for the [NOO], [NNO], or [COO]
systems.\cite{MM.no2:2020,MM.n2o:2020,MM.co2:2021} One possible reason
is in the coordinates used which was a Jacobi-system. Transforming to
hyperspherical coordinates would be a testable alternative. However,
using KerNN with distance-dependent kernels - as pursued in the
present work - yielded a highly accurate representation of the
reactive PES without too much technical difficulty. A possible future
improvement consists in transfer learning the present KerNN PES to the
FCI/aug-cc-pV5Z level of
theory.\cite{MM.tl:2022,MM.tl:2023,MM.tl:2024} It is noted that "brute-force" calculation of a sufficient
  number of reference points $(\sim 5 \times 10^4)$ at the
  FCI/aug-cc-pV5Z level even for the three-electron system is extremely time consuming or even unfeasible.\\

\noindent
The ability to run statistically significant numbers of QCT
trajectories on high-precision PESs also allows to discover more
exotic dynamics. This was highlighted for the atom exchange reaction
channel, see Figure \ref{fig:qct_traj2}, which was not the prime focus
of the present work. One of the trajectories run at 100 K features
roaming of the He-atom around the H$_2^+$ ion on a time scale of 1 ps
before the reactive collision leads to H-atom
exchange.\cite{bowman.roaming:2011,bowman.collision:2020} With decreasing temperature it is anticipated that the number of such collision events increases further.\\

\noindent
It is also of interest to note that the overall shapes of the KerNN
and cR-PES surfaces agree around the [He-H-H]$^+$ structure but differ
for the [H-He-H]$^+$ conformation, see Figures \ref{fig:pes} and
\ref{fig:1dcut}. Nevertheless, the final state probabilities and
distributions from QCT simulations agree closely, see Figure
\ref{fig:prob-kernn-log}. On the other hand, the initial
state-selected $T-$dependent rates (Figures \ref{fig:rates} and
\ref{fig:rates-j0to5}) behave differently. In other words, for
dynamically averaged properties the differences in the two PESs do not
manifest themselves (see also integral cross sections) whereas at a
state-selected level such differences can be observed. This may be
owed to the fact that cR-PES was constructed by combining accurate
two-body terms\cite{MM.heh2:2019} with a three-body term from a
different representation.\cite{ramachandran:2009} In a somewhat
different context it was already noted earlier that even within the
same representation of a PES eliminating certain terms does not
necessarily lead to a globally valid PES: starting from the global and
reactive N$_2$ + N$_2$ PES\cite{paukku:2013} a global N + N$_2$ PES
was determined by moving the fourth nitrogen atom to
infinity.\cite{mankodi:2017} Subsequently, it was demonstrated that
for certain geometries such a procedure yields a valid PES whereas for
others this is not the case.\cite{varga:2021}\\

\noindent
In conclusion, the KerNN representation of UCCSD(T)/aug-cc-pV5Z
reference data provides a globally valid description inter- and
intramolecular interactions. This is only partially true for the
cR-PES description. QCT and TIQM simulations for the ${\rm H} + {\rm
  HeH}^+(v=0,j=0) \rightarrow {\rm H}_2^+ + {\rm He}$
reaction using both PESs for determining state-selected $T-$dependent
rates -- which were shown\cite{sahoo.refined.2024} to be close to the
thermal rate for the cR-PES at the temperatures considered -- are
consistent with the $T-$independent Langevin rate $k_{\rm L} =
2.1\times 10^{-9}$ cm$^3$/molecule/s. State-averaged properties from
classical and quantum simulations using these two PESs are consistent
with one another whereas for state-selected properties, the
differences between the PESs manifest themselves in the
observables. \\

\section{Methods}

\subsection{The KerNN PES}
KerNN is a framework for learning reactive and non-reactive molecular
PESs and has recently been introduced.\cite{mm.kernn:2024} It combines
one-dimensional reproducing kernels, that have a meaningful asymptotic
form, with a small NN (hence, KerNN). While the training and
evaluation of the PES can be carried out in Python, the simplicity of
the approach permits the seamless implementation in Fortran.  This is
particularly relevant for long-time simulations which require a large
number of sequential force predictions. In the following, an overview
of the \textit{ab initio} reference data, the reproducing kernels and
the NN is given, while Reference~\citenum{mm.kernn:2024} offers a
comprehensive overview of the technical details including the
reference data.\\

\noindent
The \textit{ab initio} reference data for the H + HeH$^+$ system is
based on the UCCSD(T)/aug-cc-pV5Z level of theory and all \textit{ab
  initio} calculations were carried out using Molpro\cite{MOLPRO}.
The data set contains energies and forces for a total of 62834
conformations. These were generated on a regular grid in Jacobi
coordinates with a tighter grid for small interatomic distances (see
Table~\ref{sitab:heh2plus_grid} taken from
Ref.~\citenum{mm.kernn:2024}) and from \textit{NVT} MD simulations at
1500 K at the semi-empirical GFN2-xTB~\cite{bannwarth2019gfn2}
level. The data set was split roughly according to 80/10/10~\% into
training/validation/test data and the zero of energy was taken with
respect to the free atoms (\textit{i.e.}, [He + H + H]$^+$ is at 0
kcal/mol).\\

\noindent
For featurization of the molecular structure, one-dimensional
reciprocal power reproducing kernels were used. Such kernels were
shown to represent diatomic potential energy curves
reliably\cite{ho96:2584} and for the present work the $k^{[3,3]}$
variant
\begin{align}\label{eq:33kernel}
    k^{[3,3]}(r, r') = \frac{3}{20 r_>^4} - \frac{6}{35}
    \frac{r_<}{r_>^5} + \frac{3}{56}\frac{r_<^2}{r_>^6}
\end{align}
was employed. Here, $r_<$ and $r_>$ correspond to the smaller and
larger values of $r$, and $r$ and $r'$ are an interatomic distance of
a query structure and a reference structure (optimized, linear H-H-He
arrangement), respectively. Hence, the one-dimensional kernels
effectively serve as a similarity function between pairs of
interatomic distances $r$ and $r'$ and decay smoothly and
monotonically towards zero for large $r_i$ according to
Eq. \ref{eq:33kernel}.\\

\noindent
The three $k^{[3,3]}$ (one for each interatomic distance) are the
descriptors and serve as the input to a fully connected feed-forward
NN. The basic building blocks of NNs are dense layers
\begin{align}
    y =\sigma (\bm{Wx} + \bm{b}),
\end{align}
which are stacked and combined with a non-linear activation function
$\sigma$. In this work, soft plus activations were used throughout
except for the output layer, which was a linear transformation. The
learnable parameters of KerNN were then fitted to reference energies
and forces by minimizing a mean squared error loss using
AMSGrad~\cite{kingma2014adam}. Further details on the NN and the
training procedure can be found in
Reference~\citenum{mm.kernn:2024}.\\

\subsection{The corrected Potential Energy Surface (cR-PES)}
The corrected PES (cR-PES),\cite{sahoo.refined.2024} was developed to
eliminate an artificial energy barrier in the HeH$^+$ + H entrance
channel of the RFCI8 PES.\cite{ramachandran:2009} Depending on the
level of theory and order of the polynomial expansion the height of
this spurious barrier ranges from 5~cm$^{-1}$ to 38~cm$^{-1}$ (0.66
meV to 4.8 meV) which is particularly problematic for low-energy
recombination reactions. The RFCI8 PES was based on $\sim 1500$
reference calculations at the FCI/cc-pVQZ level of theory and
represented as a many-body expansion using Morse-type coordinates
$\rho \sim R \exp{(-\beta R)}$ where $R$ is a distance and $\beta$
mainly controls the slope for small $R$.\\

\noindent
The new global and reactive surface cR-PES\cite{sahoo.refined.2024}
was also expressed as a many-body expansion\cite{var88:255} with the
HeH$^+$ and H$_2^+$ two-body potentials including explicit and
accurate long-range interactions from the
literature\cite{MM.heh2:2019} combined with the three-body potential
of RFCI8.\cite{ramachandran:2009} Incorporation of accurate long-range
interaction terms essentially removed the artificial energy barrier in
the H + HeH$^+$ channel presenting a typical exoergic-barrierless
ion-neutral characteristic dynamics for the reaction. The topography
of the long-range region of the corrected, barrierless cR-PES was
validated by comparing with MRCI+Q/aug-cc-pV6Z {\it ab initio}
energies which confirmed the accuracy of the cR-PES for the H + HeH$^+$ channel.\\

\subsection{QCT Simulations}
The QCT simulations in the present work follow previously established
methodologies.\cite{MM.rkhs:2017,MM.co2quantum.2022,MM.no2:2020}
Therefore, only specific technical aspects are briefly summarized
here. Simulations were performed over a temperature range of 10 to
1000 K, with $5 \times 10^5$ trajectories executed to ensure
convergence of the observables at each temperature. Hamilton's
equations of motion were solved using a fourth-order Runge-Kutta
numerical method with a time step of $\Delta t = 0.05$~fs, ensuring
the conservation of total energy and angular momentum throughout the
dynamics. Initial conditions were sampled using standard Monte Carlo
methods.\cite{truhlar:1979reactive} As the associated quantum numbers
with product diatoms are real-valued, their necessary assignment to
integer values was made using histogram binning and Gaussian binning
methods.\cite{bonnet:gaussianbin,MM.cno:2018} Since both approaches
yielded similar results, only those obtained from the histogram
binning method are discussed further.\\

\noindent
Reaction cross sections and corresponding rate coefficients were
determined from following established protocols to run and analyze the
QCT simulations.\cite{MM.cno:2018,MM.n3:2024} The cross section for an
ensemble of trajectories was determined from
\begin{equation}
    \sigma_{\rm x} = \pi b_{\rm max}^2 P
\end{equation}
where
\begin{equation}
    P = \frac{N_{\rm x}}{N_{\rm tot}}.
\end{equation}
Here, $N_{\rm x}$ is the number of trajectories corresponding to the
event of interest and $N_{\rm tot}$ is the total number of
trajectories in the ensemble, and $b_{\rm max}$ is the maximum impact
parameter. The $T-$dependent rates at a particular temperature $T$ were
obtained according to
\begin{equation}
    k(T) = g(T) \sqrt{\frac{8k_{\rm B}T}{\pi \mu}} \pi b_{\rm max}^2 P 
\end{equation}
where $g(T)$ is the electronic degeneracy factor (here equal to 1),
$k_{\rm B}$ is the Boltzmann constant and $\mu$ denotes the reduced
mass of the collision system.\\

\subsection{Quantum Scattering Calculations}
The quantum scattering dynamics of the reaction was investigated using
the time independent quantum mechanical (TIQM) scattering method, as
implemented in the \texttt{ABC} program.\cite{sko00:128} This program
employs a coupled-channel hyperspherical coordinate approach to solve
the time-independent Schr\"{o}dinger equation for the nuclear motion
of a triatomic reactive system. The diatomic rovibrational wave
functions, including both open and closed helicity states of all
available reactive channels, were used to construct the
coupled-channel basis functions. The resulting set of coupled-channel
hyperradial equations are then solved using the constant reference
potential log-derivative method.\cite{M:JCP86} The scattering matrix
(S-matrix) was obtained by imposing asymptotic boundary conditions at
large hyperradius.\\

\noindent
In a single run, the S-matrix elements were computed over a specified
energy range for a given total angular quantum number $J$ and
triatomic parity eigenvalue $P$ across all three arrangement channels
(here H$_{\rm A}$ + HeH$_{\rm B}^+$ , HeH$_{\rm A}^+$ + H$_{\rm B}$,
and He + H$_2^+$ ). Next, the parity-adapted S-matrix elements were
appropriately combined to obtain the helicity-representation S-matrix
elements, $S^{J}_{vj\Omega \rightarrow v'j'\Omega{'}} (E_{\rm col})$,
by using Eqs. (1) and (2) of Ref. \citenum{sko00:128}. The symbols
$v$, $j$, and $\Omega$ denote the vibrational, rotational, and
helicity quantum numbers of the reactant channel, respectively, and
the corresponding primed quantities are those for the product
channel. Reaction observables such as cross sections and rate
coefficients were calculated in the helicity representation.\\

\noindent
The initial rovibrational state-selected cross section was calculated
according to
\begin{eqnarray}
    \sigma_{vj} (E_{\rm col}) = \frac{\pi}{\tilde{k}_{vj}^2 (2j+1)} \sum_{v{'}
      j{'}} \sum_{J=0}^{J_{\text{max}}} \sum_{\Omega \Omega{'}}
    g_{j{'}} (2J+1) \big| S^{J}_{vj\Omega \rightarrow v'j'\Omega{'}}
    (E_{\rm col}) \big|^2 .
    \label{eq:s2sics}
\end{eqnarray}
Here, $\tilde{k}_{vj}$=$\sqrt{2\mu E_{\text{col}}}/\hbar$, with $\mu$
being the atom-diatom reduced mass of the reactant channel, and
$E_{\rm col}$ being the collision energy which is the total energy
minus the rovibrational energy of the reactant diatom. The degeneracy
factor $g_{j{'}}$ for the H$_{2}^{+}$ diatom is $g_{j{'}} = 3/2$ for
odd and $g_{j{'}} = 1/2$ for even $j'$ quantum
number. \cite{Chao_JCP_117_8341_2002, Zhang_Miller_theta_JCP_1989} \\

\noindent
Formally, the initial state-selected $T-$dependent rate coefficients
are related to the corresponding cross sections ($\sigma_{vj}$) by
thermal averaging over a Maxwell-Boltzmann velocity distribution
according to
\begin{equation}
    k_{vj} (T) = \sqrt{\frac{8k_{B}T}{\pi \mu}} \frac{1}{(k_{B}T)^{2}}
    \int_{E_{\rm col}^{\rm min}}^{\infty} E_{\text{col}} \sigma_{vj} (E_{\text{col}})
    e^{-E_{\text{col}}/k_{B}T} \text{d}E_{\text{col}}  .
    \label{eq:rateconstant}
\end{equation}
In practice, the Maxwell-Boltzmann translational energy distribution
decays at higher $E_{\rm col}$. Hence, the upper integration limit
$E_{\rm col}^{\rm max} \neq \infty$ but depends on the
temperature. Both values, $E_{\rm col}^{\rm min}$ and $E_{\rm
  col}^{\rm max}$ were chosen such as to converge the rate
coefficients. The numerical integration was performed here by
trapezoidal rule and the converged numerical parameters used in the
\texttt{ABC} code are given in Table \ref{sitab:parameter1}.\\

\begin{figure}[H]
    \centering
    \includegraphics[width=0.5\linewidth]{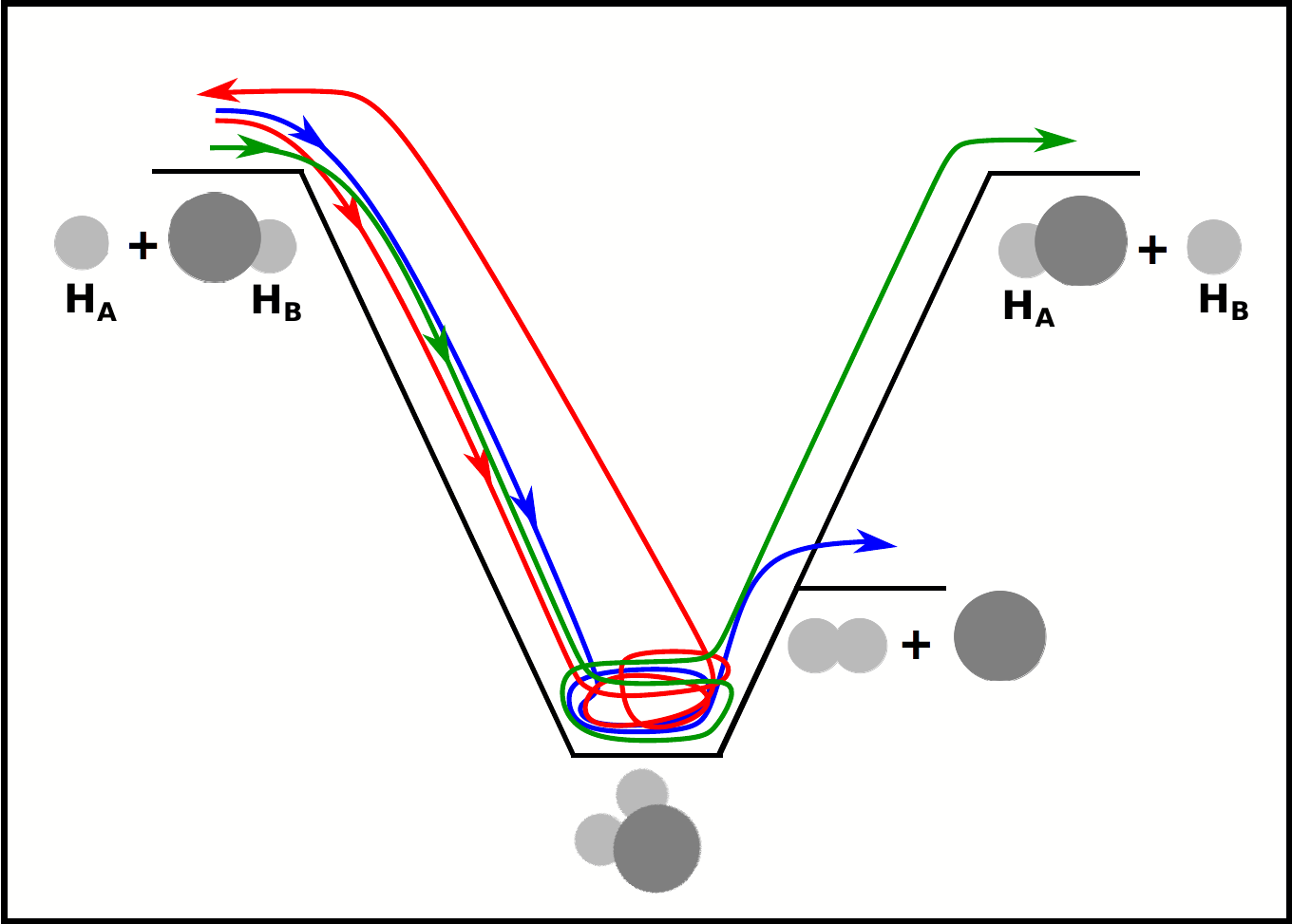}
    \caption{Graphical table of content.}
    \label{fig:enter-label}
\end{figure}

\section*{Acknowledgment}
The authors gratefully acknowledge financial support from the Swiss
National Science Foundation through grants $200020\_219779$ (MM),
$200021\_215088$ (MM), the NCCR-MUST (MM), the AFOSR (MM), and the
University of Basel. This article/publication is based upon work from
COST Action COSY CA21101, supported by COST (European Cooperation in
Science and Technology) (to MM). This work has also been supported by
the Agence Nationale de la Recherche (ANR-HYTRAJ) under Contract
No. ANR-19-CE30-0039-01 (to YS). YS and JS acknowledge the support
from the High Performance Computing Platform MESO@LR at the University
of Montpellier. We thank D. Talbi for correspondence on details of
some of the electronic structure calculations.\\

\clearpage

\renewcommand{\thetable}{S\arabic{table}}
\renewcommand{\thefigure}{S\arabic{figure}}
\renewcommand{\thesection}{S\arabic{section}}
\renewcommand{\d}{\text{d}}
\setcounter{figure}{0}  
\setcounter{section}{0}  
\setcounter{table}{0}

\newpage

\noindent
{\bf SUPPORTING INFORMATION: Reaction Dynamics of the H + HeH$^+$
  $\rightarrow$ He + H$_2^+$ System}

\begin{table}[H]
\begin{tabular}{ccc}\hline
Quantity / Diatom     & H$_2^+$ & HeH$^+$\\\hline
$r$ & 0.5/5/0.2 and 6/50/1 & 0.5/5/0.2\\
$R$ & 0.5/5/0.2 and 6/50/1 & 0.5/5/0.2\\
$\Theta$ & 0/180/15 & 0/180/15\\
$N_{\rm tot}$ & 58357 & 6877\\\hline
\end{tabular}
\caption{Grid for the triatomic HeH$_2^+$ potential. All values are
  given in \AA\/ and degrees, respectively. The quantities are given
  as min/max/step.}
\label{sitab:heh2plus_grid}
\end{table}

\begin{table}[htbp]
\centering
\caption{Converged values of the numerical parameters used in the
  \texttt{ABC} calculations.  ( \textit{jmax}: maximum rotational
  quantum number considered in any channel, \textit{emax}: the maximum
  total energy below which all the rovibrational levels, both open and
  closed, are included in the basis set, \textit{rmax}: maximum value
  of the hyperradius, \textit{mtr}: number of log derivative
  propagation sectors, \textit{dnrg}: scattering energy increment,
  $J_{\text{max}}$: maximum value of the total angular quantum number
  required to converge the cross section within the specified energy
  range, and \textit{kmax}: the helicity truncation parameter.)}
\setlength{\tabcolsep}{4.6pt} \renewcommand{\arraystretch}{1.4}
\begin{tabular}{cccccccccc}
\hline
{$E_{\rm col}$} range (eV) & \textit{jmax} & \textit{emax} (eV) &  \textit{rmax} (a$_0$) & \textit{mtr} & \textit{dnrg} (eV) & $J_{\text{max}}$ & \textit{kmax} \\
\hline \hline 
 1.0E-5 $-$ 1.0E-4 & 24 & 1.6 & 200 & 4976 & 2.0E$-$6 & 7 & 7 \\
 1.0E-4 $-$ 1.0E-3 & 24 & 1.6 & 160 & 3976 & 2.0E$-$5 & 10 & 7 \\
 1.0E-3 $-$ 1.0E-2 & 24 & 1.6 & 70 & 1726 & 2.0E$-$4 & 16 & 7 \\
 0.01 $-$ 0.05 & 24 & 1.6 & 50 & 818 & 0.001 & 26 & 9 \\
 0.05 $-$ 0.1 & 24 & 1.6 & 35 & 568 & 0.001 & 33 & 11 \\
 0.1 $-$ 0.3 & 24 & 1.6 & 26 & 250 & 0.005 & 45 & 11 \\
 0.3 $-$ 0.5 & 26 & 1.7 & 26 & 250 & 0.005 & 53 & 11 \\
 0.5 $-$ 1.0 & 30 & 1.8 & 26 & 250 & 0.02 & 67 & 12 \\
\hline 
\end{tabular}
\label{sitab:parameter1}
\end{table}

\begin{figure}[H]
    \centering
    \includegraphics[width=0.8\linewidth]{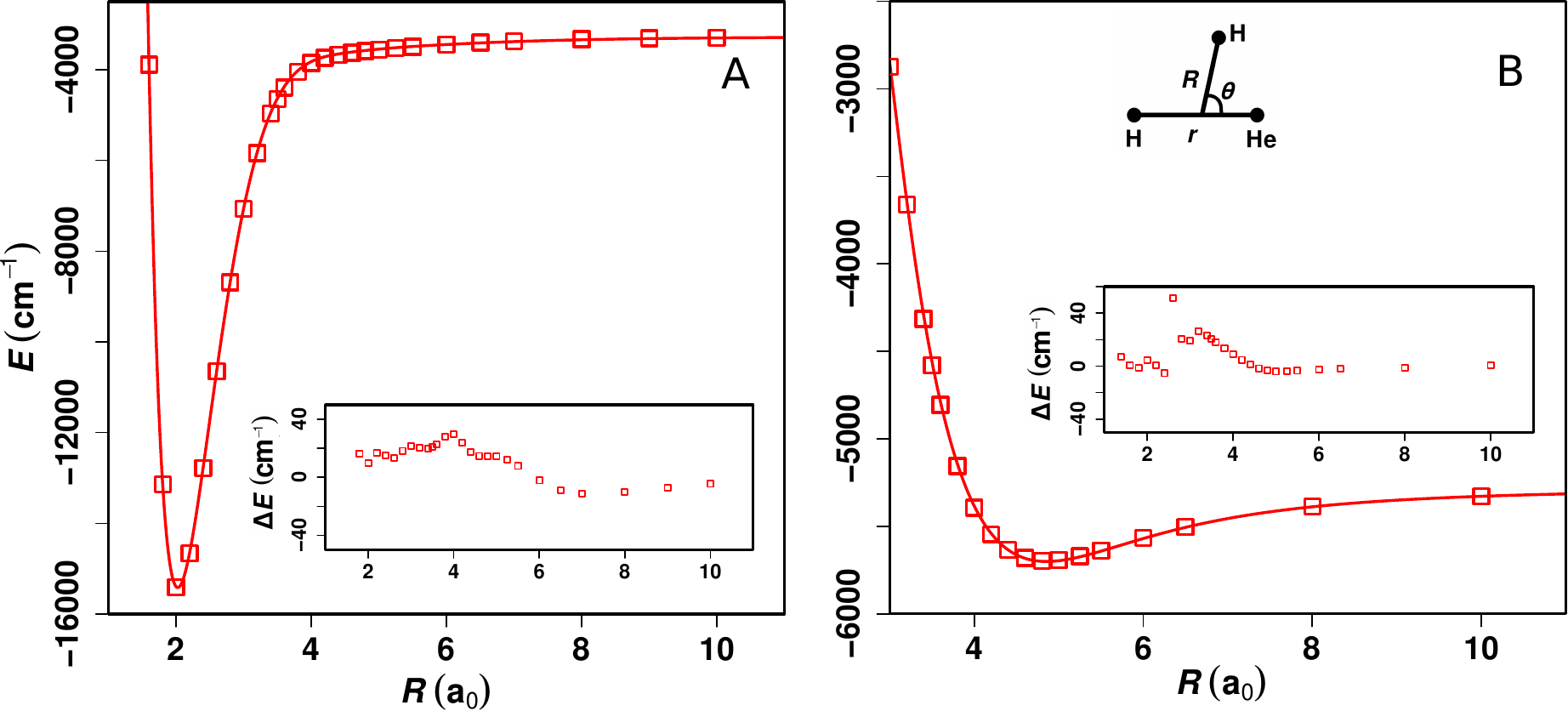}
    \caption{One-dimensional cuts comparing FCI/aug-cc-pV5Z \textit{ab
        initio} energies (squares) with the KerNN PES representing
      UCCSD(T)/aug-cc-pV5Z reference data (solid lines). Panel A:
      $\theta = 20.40^\circ$, $r = 3.0$ a$_0$ and Panel B: $\theta =
      78.393^\circ$, $r = 2.65$ a$_0$. Insets report $\Delta E =
      E_{\rm KerNN} - E_{\rm FCI}$ for the respective geometries.}
    \label{sifig:pes-valid}
\end{figure}

\begin{figure}[H]
    \centering
    \includegraphics[width=10.5cm]{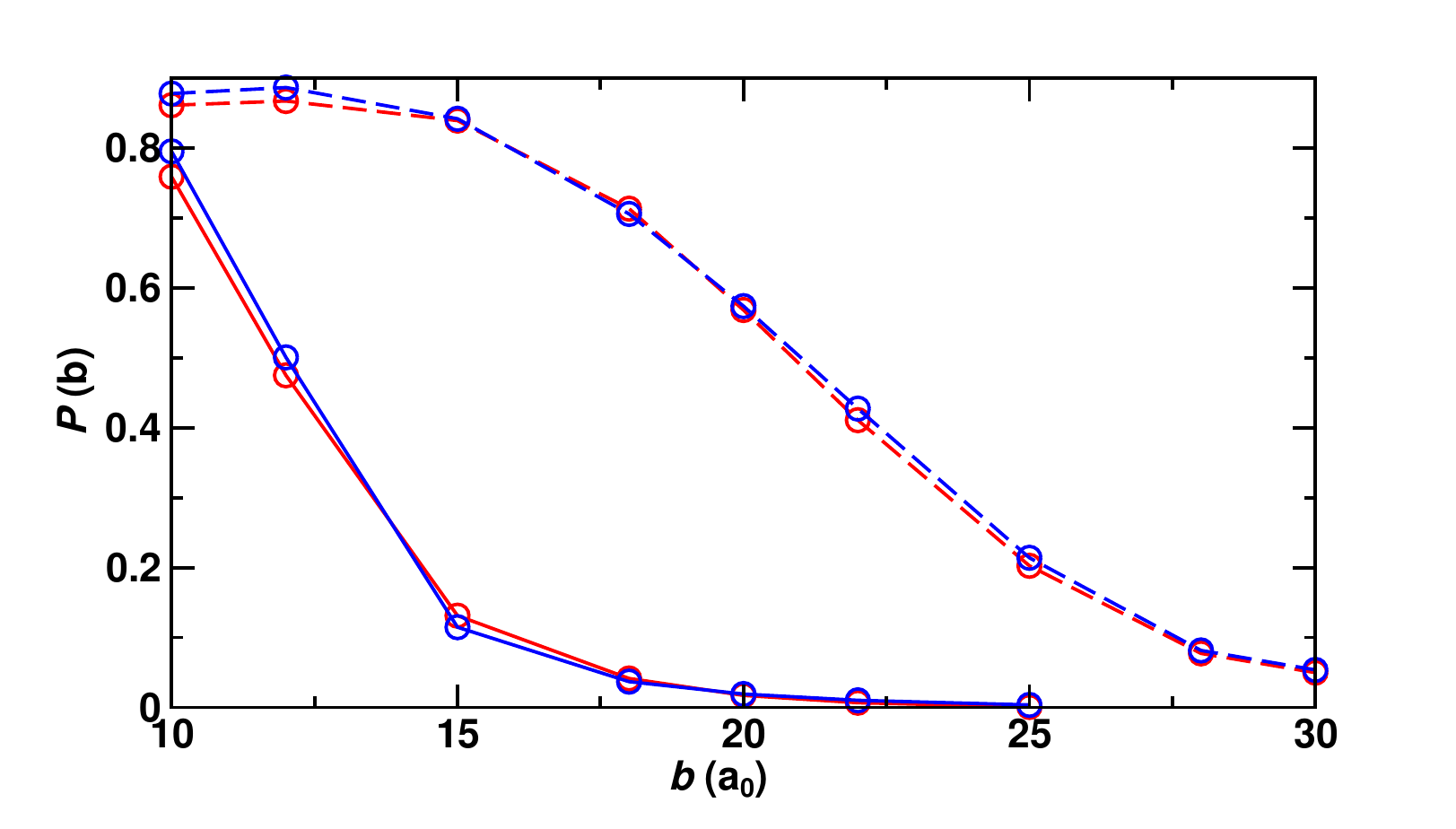}
    \caption{The opacity function from QCT for KerNN (red) and the
      cR-PES (blue) for H + HeH$^+$($v=0, j=0$) $\longrightarrow$ He +
      H$_2^+$ reaction for $T=10$ K (dashed line) and $100$ K (solid
      line) and $r_{\rm 0}= 30$ a$_0$.}
    \label{sifig:opacity}
\end{figure}

\begin{figure}
    \centering
    \includegraphics[width=0.75\linewidth]{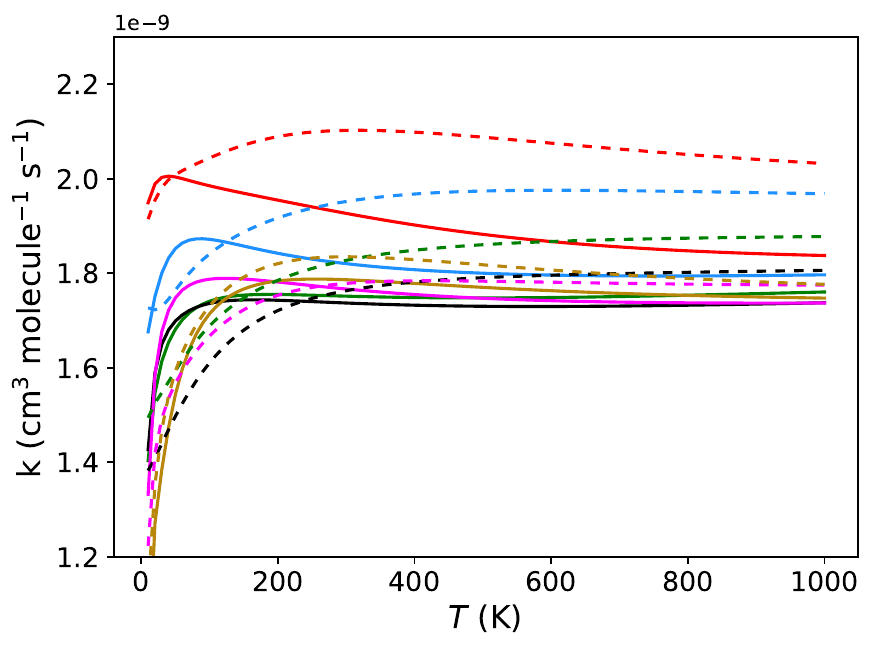}
    \caption{Initial-state-selected rate coefficients $k_{v,j}(T)$
      from TIQM simulations for the ${\rm H} + {\rm HeH}^+(v=0,j) \rightarrow {\rm He} + {\rm H}_2^+$ reaction.  Color code for
      the initial states of HeH$^+$: $j=0$ (red), $j=1$ (blue), $j=2$
      (green), $j=3$ (black), $j=4$ (magenta), and $j=5$
      (brown). Solid and dashed lines correspond to simulations using
      the KerNN and cR-PES surfaces, respectively. }
    \label{sifig:tiqm-rates-j0j5}
\end{figure}

\begin{figure}
    \centering
    \includegraphics[width=0.9\linewidth]{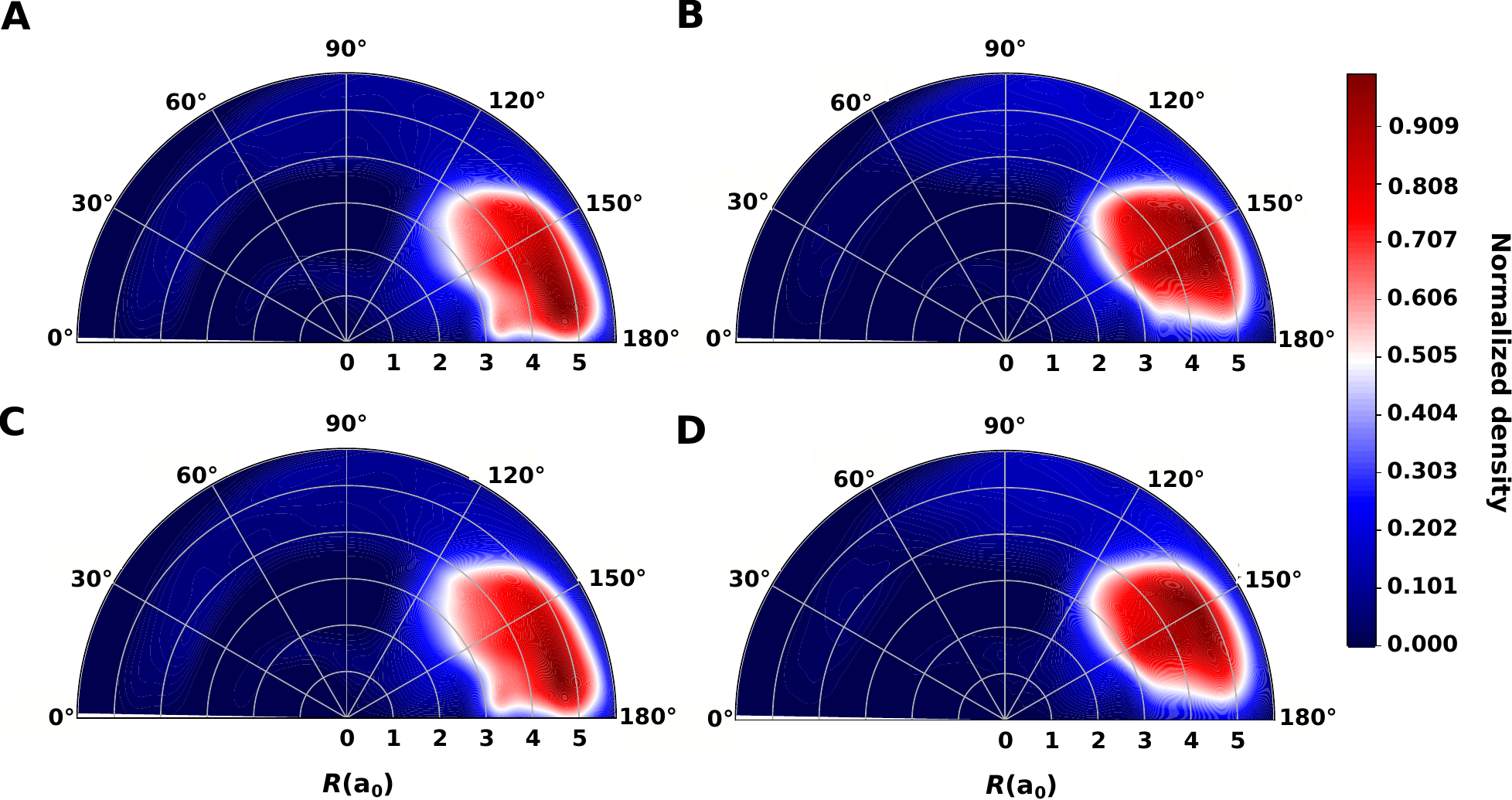}
    \caption{Geometrical sampling of incoming ``H" around the center
      of mass of He--H$^+$ during complex formation using 5000
      trajectories carried out with KerNN (Panel A and B) and cR-PES
      (Panel C and D) for $T=$ 100 K (left-hand side) and 1000 K
      (right-hand side). Here $\theta = 0^\circ$ corresponds to
      [H-He-H]$^+$ and $\theta = 180^\circ$ corresponds to the
      [He-H-H]$^+$ conformation.}
    \label{sifig:complex-geo}
\end{figure}

\begin{figure}[H]
    \centering
    \includegraphics[width=0.5\linewidth]{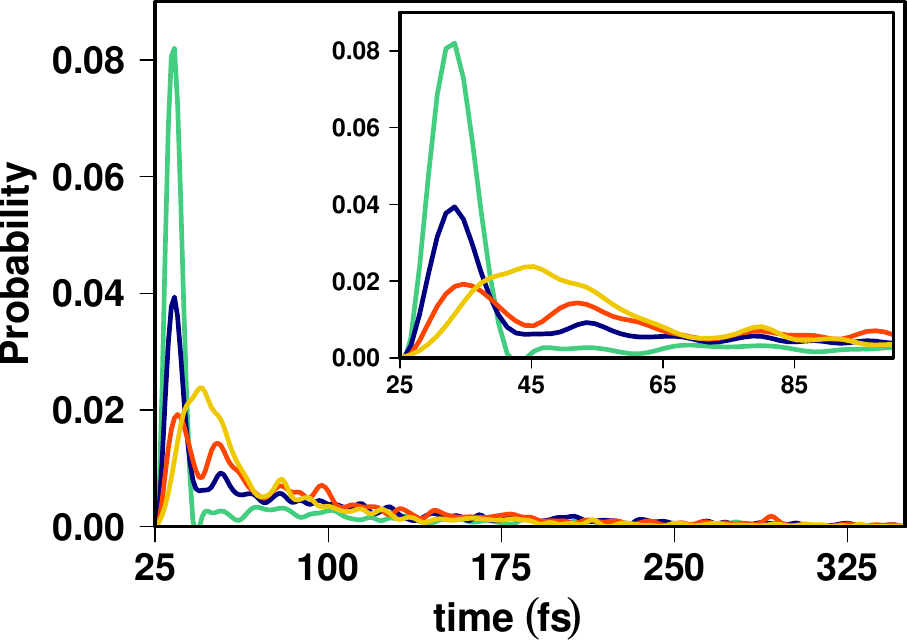}
    \caption{Normalized lifetime distributions $P(\tau)$ of the
      collision complex for the H + HeH$^+$($v=0, j=0$)
      $\longrightarrow$ He + H$_2^+$($v',j'$) reaction from QCT
      simulations using KerNN. For 10 K (green), 100 K (navyblue), 300
      K (red), and 1000 K (yellow), 2000 independent QCT simulations
      were analyzed. The lifetime of the collision complex is defined
      as the time elapsed from the first instance when each
      intermolecular separation is smaller than 6 a$_0$ until the
      moment any one of the separations is larger than 6 a$_0$.}
    \label{sifig:lifetime}
\end{figure}

\begin{figure}[H]
  \centering \includegraphics[width=16.5cm]{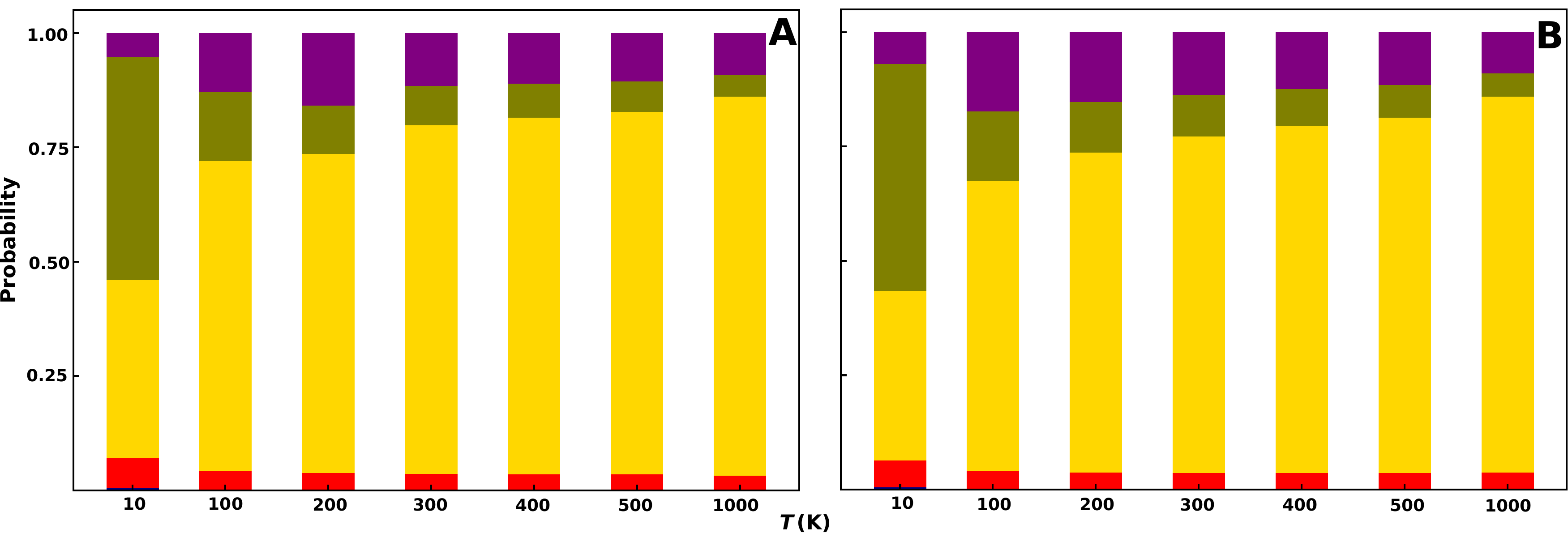}
    \caption{Relative fraction of the different possible reaction
      channels on a linear $y-$scale using KerNN (left) and cR-PES
      (right) depending on temperature. See also Figure
      \ref{fig:prob-kernn-log}.}
\label{sifig:prob-kernn-linear}
\end{figure}

\begin{figure}[H]
    \centering
    \includegraphics[width=0.55\linewidth]{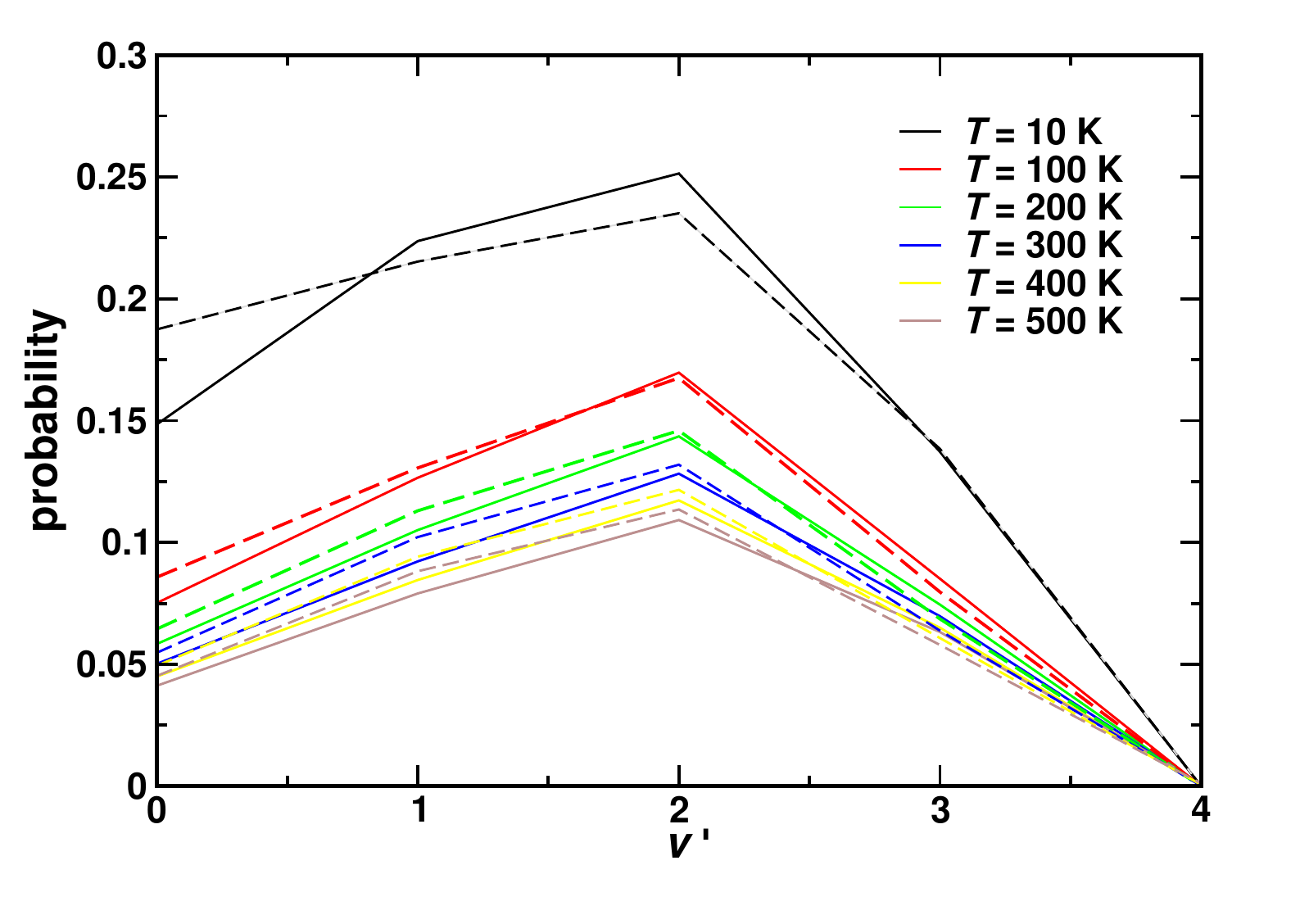}
    \caption{Product diatom vibrational level distributions at
      different temperatures for H + HeH$^+$($v=0, j=0$)
      $\longrightarrow$ He + H$_2^+$($v',j'$) reaction using KerNN
      (solid lines) and cR-PES (dashed lines).}
    \label{sifig:1d-pv}
\end{figure}

\begin{figure}[H]
    \centering
    \includegraphics[width=0.6\linewidth]{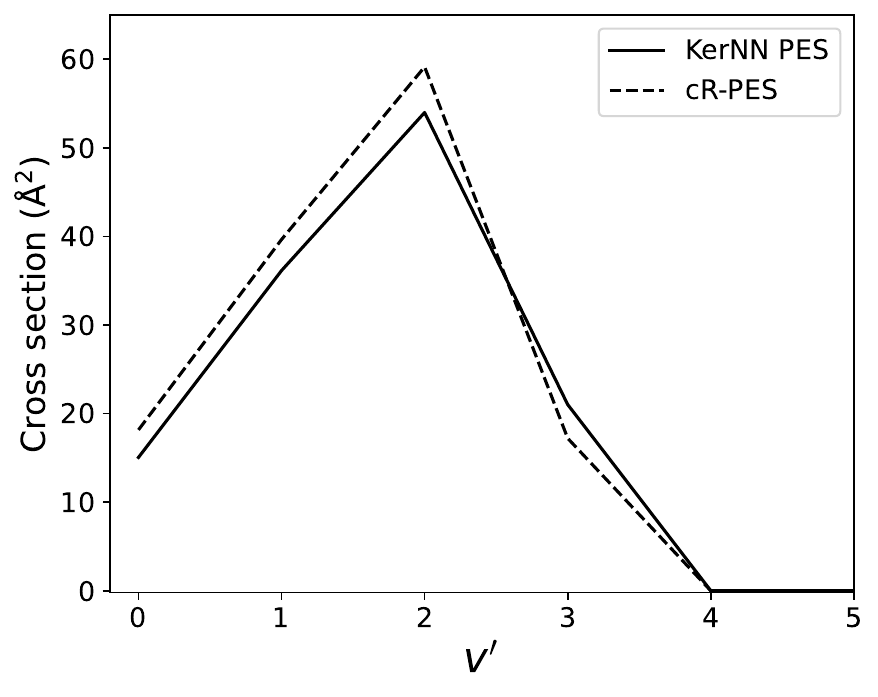}
    \caption{Product diatom vibrational level distributions at
      $E_{col}$ = 0.01 eV ($\approx$ 116 K) from TIQM simulations for
      H + HeH$^+$($v=0, j=0$) $\longrightarrow$ He + H$_2^+$($v',j'$)
      reaction using the KerNN and cR-PES surfaces. The collision
      energy in TIQM simulation is fixed whereas for QCT simulation it
      is drawn from Maxwell-Boltzmann distribution at temperature
      $T$. }
    \label{sifig:pv-quantum}
\end{figure}

\section{Total Kinetic energy release}
The computed TKER is the sum of the He and H$_2^+$ fragments
translational energy contributions $E_{trans,\alpha} = \Vec{p}_{\rm
  CM,\alpha} ^ 2$$/2M_{\rm \alpha}$ where $M_{\rm \alpha}$ is the mass
and $\vec{p}_{\rm CM,\alpha}$ is the center of mass momentum of
fragments $\alpha$ obtained as the sum of the respective atom
momenta. In QCT simulations, the final $v'$ and $j'$ quantum numbers
are real-valued rather than strictly quantized. Therefore, a rigorous
selection criterion was applied to extract trajectories for
analysis. Only trajectories for which the (real-valued) final $v'-$
and $j'-$values differ by no more than 0.01 from the nearest quantized
values were considered. Out of the $5 \times 10^5$ trajectories that
were run, only 0.1\% met this criterion and were used for TKER
analysis.\\

\newpage

\bibliography{refs}

\end{document}